\documentclass[11pt]{iopart}

\usepackage{graphicx}
\usepackage{subfigure}
\begin{document}

\title[Solution of the time-fractional Gross-Pitaevskii equation]{Solutions of the Gross-Pitaevskii and
time-fractional Gross-Pitaevskii equations for different potentials
with Homotopy Perturbation Method}

\author{Neslihan \"{U}zar, S. Deniz Han, Tu\~{g}ba T\"{u}fek\c{c}i and Ekrem Ayd{\i}ner}

\address{\.{I}stanbul University Theoretical Physics Research Group,
\.{I}stanbul University, Department of Physics Tr-34134
\.{I}stanbul, Turkey} \ead{ekrem.aydiner@istanbul.edu.tr}

\begin{abstract}
In this study, after we have briefly introduced the standard
Gross-Pitaevskii equation, we have suggested fractional
Gross-Pitaevskii equations to investigate the time-dependent ground
state dynamics of the Bose-Einstein condensation of weakly
interacting bosonic particle system which can includes non-Markovian
processes or non-Gaussian distributions and long-range interactions.
Only we focused the time-fractional Gross-Pitaevskii equation and
have obtained solutions of the standard Gross-Pitaevskii and
time-fractional Gross-Pitaevskii equations for attractive and
repulsive interactions in the case external trap potentials $V(x)=0$
and optical lattice potential $V\left( x\right) =\pm\sin^{2}x$ by
using Homotopy Perturbation Method. We have found that the Homotopy
Perturbation Method solutions of the Gross-Pitaevskii equation for
these potentials and interactions are the same analytical results of
it. Furthermore we have also found that solutions of the
time-fractional Gross-Pitaevskii equation for these potentials and
interactions can be given in terms of Mittag-Leffler function. The
solutions of the time-fractional Gross-Pitaevskii equation provide
that the time evolution of the ground state dynamics of
Bose-Einstein condensation of bosonic particles deviates exponential
form, and evolutes with time as stretched exponentially.
\end{abstract}

\maketitle

\section{Introduction}

The zero temperature Bose-Einstein condensation (BEC) of dilute or
weakly interacting bosons are generally described by
Gross-Pitaevskii equation which is s a self-consistent mean field
nonlinear Schr\"{o}dinger equation \cite{gp1Gross,gp2Pitaevskii}. It
is clearly known that this equation does not include memory effects
of the non-Markovian processes, non-Gaussian distribution of
particles, the long-range inter-particle interaction effects or the
fractal structure of the interacting space between particles in
condensate phase. However, it is shown that some experimental
observations in a real real dilute or weakly interacting bosonic
systems deviates from theoretical predictions i.e. some experimental
results do not comply with theoretical calculations. For example,
for dilute atomic gases that involve interacting atoms or molecules
such as $^{87}$Rb, $^{23}$Na and $^{7}$Li which are trapped in a
harmonic oscillator potential, there are significant differences
between the theoretical and experimental transition temperatures in
Bose-Einstein condensation. Similarly, for $^{4}$He, the
experimentally observed value of the transition temperature to BEC
is $T_c=2.17$ K, whereas the calculated one is $T_c=3.10$ K in
conventional Bose-Einstein thermostatistics
\cite{Pethick,Anderson,Jin,Davis,Mewes,Bradley}. These contractions
between experimental observation and theoretical predictions shows
that the memory effects of the non-Markovian processes, non-Gaussian
distribution of particles, the long-range inter-particle interaction
effects or the fractal structure of the interacting space in real
bosonic systems have a predominant role below the critical
temperature $T_{c}$ \cite{ekrem}. For this reason, we can claim that
this standard approaches remain insufficient in the investigation
ground state dynamics of the physical behavior of weakly interacting
quantum boson gases at low temperatures i.e. below the $T_{c}$. At
finite temperature below the $T_{c}$ and even though at zero
temperature, the non-Markovian processes, non-Gaussian distribution
of particles, the long-range effects between particles or the
fractal structure of the interacting space in the Bose-Einstein
condensate phase must be taken into account in the model. However,
to the best of our knowledge on the literature, there does not exist
a sound mathematical basis for these efforts in the literature.

In this study, we will adopt fractional calculus, which involves
fractional derivatives and integrals. With the help of fractional
calculus, it is possible to take into account memory effects of the
non-Markovian processes, non-Gaussian distribution of particles, the
long-range inter-particle interaction effects or the fractal
structure of the interacting space \cite{ekrem,Tarasov1,Tarasov2}.
Because in the fractional calculus concept, these unusual effects
that are disappearing in the Markovian process can be represented
the space-fractional and time-fractional derivative operators
\cite{Laskin1,Naber1}. Therefore, here we will define fractional
Gross-Pitaevskii equations, and we will focus only time-fractional
Gross-Pitaevskii equation to investigate ground state dynamics of
the Bose-Einstein condensate of dilute or weakly interacting boson
gas has repulsive or attractive interactions for zero external
potential and optical lattice potentials at zero temperature.

As it is known that time-dependent Gross-Pitaevskii equation is a
nonlinear partial differential equation and it is difficult to find
analytical solution of this type of equation for especially complex
potentials. Therefore, the numerical solution methods are generally
used to find solution of them. In order to solve fractional ordinary
differential equation, integral equation and fractional partial
differential equations in physics and other areas of science,
several analytical and numerical methods have been proposed. The
most commonly used ones are; Adomain Decomposition Method (ADM)
\cite{a1Momani,a2Momani,a3Odibat,a4Momani,a5Momani,a6Momani},
Variational Iteration Method (VIM)
\cite{a4Momani,a5Momani,a6Momani,v1He,v2Odibat}, Fractional
Difference Method (FDM) \cite{f1Podlubny}, Differential Transform
Method (DTM) \cite{d1Arikoglu}, Homotopy perturbation Method (HPM)
\cite{h1Odibat,h2Monami}. Also there are some classical solution
techniques e.g. Laplace transform method, Fractional Green's
function method, Mellin transform method and method of orthogonal
polynomials \cite{f1Podlubny}. In this study, we will use the HPM
method to solve fractional Gross-Pitaevskii equations. HPM has been
proposed by He \cite{hpm1He,hpm2He} to solve linear and nonlinear
differential and integral equations. This method, which is a
coupling of the traditional perturbation method and homotopy in
topology, deform continuously to simple problem which easily solved.
The HPM can be easily applied to Volterra's integro-differential
equation \cite{vol1El}, to nonlinear oscillators \cite{os1He},
bifurcation of nonlinear problems \cite{bif1He}, bifurcation of
delay-differential equations \cite{bifd1He}, nonlinear wave
equations \cite{non1He}, boundary value problems \cite{bo1He},
quadratic Ricatti differential equation of fractional order
\cite{h1Odibat} and to other fields
\cite{o1He,o2He,o3He,o4He,o5Siddiqui,o6Siddiqui,o7He,o8Abbasbandy,o9Abbasbandy}.
This HPM yield very convergence of the solution series in most
cases, usually only a few iteration leading to very accurate
solutions.

This paper is organized as follows. In Section 2 we briefly review
Gross-Pitaevskii equation. In Section 3, we present the stationary
solution of the  Gross-Pitaevskii equation. In Section 4, we define
Gross-Pitaevskii equation with fractional orders. In Section 5, we
summarize basic definition of the fractional calculus. In Section 6,
the Homotopy perturbation method is introduced. In Section 7, we
discuss several example. Finally Section 8 is devoted to
conclusions.

\section{Gross-Pitaevskii equation}

The properties of a Bose-Einstein condensate at zero temperature are
well described by the macroscopic wave-function $\psi \left(
x,t\right)$ whose evolution is governed by the Gross-Pitaevskii
equation \cite{gp1Gross,gp2Pitaevskii} which is a self-consistent
mean field nonlinear Schr\"{o}dinger equation:
\begin{equation}
i\hbar \frac{\partial }{\partial t}\psi \left( x,t\right) =\left( -\frac{%
\hbar ^{2}}{2m}\nabla ^{2}+V\left( x\right) +g\left\vert \psi \left(
x,t\right) ^{2}\right\vert \right) \psi \left( x,t\right) \ ,  \quad
x\in \Omega \subseteq  R^{d}
\end{equation}
\begin{equation}
\psi \left( x,t\right) =0 \ , \quad x\in \Gamma =\partial \Omega,
\end{equation}
\begin{equation}
\psi \left( x,0\right) =\psi _{0}\left( x\right) \ , \quad x\in
\Omega;
\end{equation}
where $g$, $m$ and $V\left( x\right)$ are the interacting parameter
between particles, mass of the particles and external potential
applying to the particle systems, respectively. The interacting
parameter i.e. coupling constant $g$ is defined as $g=4\pi\hbar
a_{s}/m$ where $a_{s}$ the scattering length of two interacting
bosons. Coupling constant determines the interaction types between
particles. For $g<0$ and $g>0$ interactions between bosonic
particles are attractive and repulsive, respectively.

It is known that there are two important invariant of Eq.\,(1) which
are the normalization of the wave-function
\begin{equation}
N\left( \psi \right) =\int_{\Omega }\left\vert \psi \left(
x,t\right) \right\vert ^{2}dx\equiv N\left( \psi _{0}\right)
=\int_{\Omega }\left\vert \psi _{0}\left( x\right) \right\vert
^{2}dx=1 \ , \quad t\geq 0
\end{equation}
and the energy functional for $t\geq 0$
\begin{equation}
E\left( \psi \right) =\int_{\Omega }\left[ \frac{1}{2}\left\vert
\nabla \psi\left( x,t\right) \right\vert ^{2}+V\left( x\right)
\left\vert \psi\left(
x,t\right)\right\vert ^{2}+%
\frac{g}{2}\left\vert \psi\left( x,t\right)\right\vert ^{4}\right]
dx\equiv E\left( \psi _{0}\right)
\end{equation}
where $N\left(\psi\right)$ is particle number and $E\left(\psi
\right)$ is the energy of the particle systems in condensate phase.
It can be seen from Eq.\,(5) the energy functional $E\left( \psi
\right)$ consist of the three parts i.e., kinetic energy, potential
energy and interacting energy.

The magnitude square of the eigenfunction, $\left\vert \psi \left(
x,t\right) \right\vert ^{2}$, represents the probability density of
finding a particle at position $x$ and time $t$. Stationary state of
the Bose-Einstein condensed system is the independent of the time.
To find stationary solution of Eq.\,(1), we write the wave function
as
\begin{equation}
\psi \left( x,t\right) =\psi \left( x\right) e^{-i\mu t/\hbar}
\end{equation}
where $\mu$ is the chemical potential of the condensate and $\psi
\left( x\right)$ is a function independent of time. Substituting (6)
into (1) yield the equation
\begin{equation}
\mu \psi \left( x\right) =\left( -\frac{1}{2}\nabla ^{2}+V\left(
x\right) +g\left\vert \psi \left( x\right) ^{2}\right\vert \right)
\psi \left( x\right) \ , \quad x\in \Omega \subseteq R ^{d}
\end{equation}
for $\psi \left( x\right)$ under the normalization condition
\begin{equation}
\left\Vert \psi \right\Vert =\int_{\Omega }\left\vert \psi \left(
x\right) \right\vert ^{2}dx=1
\end{equation}
On the other hand, any eigenvalue $\mu$ can be computed from its
corresponding eigenfunction $\psi$ by
\begin{equation}
\mu =\mu \left( \psi \right) =\int_{\Omega }\left[
\frac{1}{2}\left\vert \nabla \psi\left( x\right) \right\vert
^{2}+V\left( x\right) \left\vert \psi\left( x\right) \right\vert
^{2}+g\left\vert \psi\left( x\right) \right\vert ^{4}\right] dx \ .
\end{equation}

\section{Soliton or stationary solutions of the Gross-Pitaevskii equation}
The most simple solution of the Eq.\,(1) is the stationary solution
of its for $V\left( x\right)=0$. When $d=1$, $V\left( x\right)=0$
the soliton solutions can be obtained (See \cite{gp2Pitaevskii}).
For repulsive interactions i.e. $g>0$ the solution of Eq.\,(1) is
given by
\begin{equation}
\psi _{d}\left( x\right) =\psi _{d}\left( x,0\right)=\sqrt{\frac{\mu }{g}}\tan h\left( \sqrt{\mu }%
x\right)
\end{equation}
This solution is known as \emph{dark solution}. Hence we can write
the time-dependent dark soliton as
\begin{equation}
\psi _{d}\left( x,t\right) =\sqrt{\frac{\mu }{g}}\tan h\left( \sqrt{\mu }%
x\right) e^{-i\mu t/\hbar}
\end{equation}
which corresponds to a local minimum of the density distribution
$\left\vert \psi \left( x,t\right) \right\vert ^{2}$. On the other
hand, for attractive interactions i.e. $g<0$, the solution of the
Eq.\,(1) is given by
\begin{equation}
\psi _{b}\left( x\right) =\psi _{b}\left( x,0\right) =\sqrt{\frac{\mu }{g}}\sec h\left( \sqrt{\mu }%
x\right) \ .
\end{equation}
This solution is known as \emph{bright solution}. Similarly,
time-dependent bright soliton solution is given by
\begin{equation}
\psi _{b}\left( x,t\right) =\sqrt{\frac{\mu }{g}}\sec h\left( \sqrt{\mu }%
x\right) e^{-i\mu t/\hbar} \
\end{equation}
which corresponding to a local maximum of the density distribution
$\left\vert \psi \left( x,t\right) \right\vert ^{2}$. These
solutions can be generalized easily in the progressive wave form.
Generalized dark soliton solution of Eq.\,(1) can be written in the
one-dimensional form
\begin{equation}
\psi \left( x,t\right) _{d}=\sqrt{\frac{\mu }{g}} \tan h\left[
\sqrt{\mu } \left( x-vt-x_{0}\right) \right]  e^{-i\mu t/\hbar}
\end{equation}
and similarly generalized bright solution solution of Eq.\,(1) is
given by
\begin{equation}
\psi \left( x,t\right) _{b}=\sqrt{\frac{\mu }{g}} \sec h\left[
\sqrt{\mu }\left( x-vt-x_{0}\right) \right]  e^{-i\mu t/\hbar}
\end{equation}
where $v$ is the speed of the soliton. Also $x_{0}$ and $t_{0}$ are
respectively initial position and time. Solitons are quantized
vertex in Bose-Einstein condensate system, which can be observation
experimentally. The shape of the solitons are change depending on
external potential. We here discussed only $V\left( x\right)=0$ case
for simplicity.

\section{Gross-Pitaevskii equations with fractional derivatives}

We remark in introduction that the memory effects of the
non-Markovian processes, non-Gaussian distribution of particles, the
long-range inter-particle interaction effects or the fractal
structure of the interacting in a bosonic system below the
condensate temperature $T_{c}$ space can leads to fractional
dynamics. Simply we say that all non-Markovian stochastic processes
of the particles or waves in condensate phase of bosonic system
generate the time-fractional Gross-Pitaevskii equation ignoring
boundary conditions and sources. Similarly, non-Gaussian
distribution of the particles or waves in condensate phase leads to
space-fractional Gross-Pitaevskii equation. On the other hand the
combination of the non-Markovian and non-Gaussian behavior yield
time and space-fractional Gross-Pitaevskii equation. These equation
are defined in Ref.\,\cite{ekrem}. In the light of these knowledge,
the one-dimensional time-fractional Gross-Pitaevskii equation can be
written in the form
\begin{equation}
i\hbar \psi \left( x,t\right) =\ _{0}D_{t}^{1-\alpha }\left( -\frac{\hbar ^{2}}{2m}%
\nabla ^{2}+V\left( x\right) +g\left\vert \psi \left( x,t\right)
^{2}\right\vert \right) \psi \left( x,t\right)
\end{equation}
where $_{0}D_{t}^{1-\alpha }$ is the Riemann-Liouville fractional
integral operator. Similarly the one-dimensional space-fractional
Gross-Pitaevskii equation can be defined as
\begin{equation}
i\hbar \frac{\partial }{\partial t}\psi \left( x,t\right) =\left( -\frac{%
\hbar ^{2}}{2m}\nabla ^{\beta }+V\left( x\right) +g\left\vert \psi
\left( x,t\right) ^{2}\right\vert \right) \psi \left( x,t\right) \ .
\end{equation}
Finally we can write time and space-fractional Gross-Pitaevskii
equation in the one-dimensional form
\begin{equation}
i\hbar \psi \left( x,t\right) =\ _{0}D_{t}^{1-\alpha }\left( -\frac{\hbar ^{2}}{2m}%
\nabla ^{\beta }+V\left( x\right) +g\left\vert \psi \left(
x,t\right) ^{2}\right\vert \right) \psi \left( x,t\right) \ .
\end{equation}
Here we focus solutions of the time-fractional Gross-Pitaevskii
equation for different potentials. But we will consider other cases
of these equations for different potential in the another study.

\section{Basic definitions for fractional calculus}

Before we discuss solution of the time-fractional Gross-Pitaevskii
equation (16) we will give basic rules for calculating fractional
differential equations in this section. Some definitions for
fractional calculus are given below
\cite{f1Podlubny,fc1Oldham,fc2Miller,fc3Samko,fc4Kilbas,fc5Baleanu,fc6Lakshmikantham}.

\paragraph{\textbf{Definition 1}} The Riemann-Liouville fractional integral
operator of order $\alpha\geq0$ of the function $f(t)\in C_{\mu}$ is
defined as for $\mu\geq-1$
\begin{equation}
_{t_{0}}D_{t}^{-\alpha }f\left( t\right) =\frac{1}{\Gamma \left(
\alpha
\right) }\int_{t_{0}}^{t}dt^{\prime }\frac{f\left( t^{\prime }\right) }{%
\left( t-t^{\prime }\right) ^{1-\alpha }}, \ \quad\alpha >0,  \
\quad t>0
\end{equation}
where $_{t_{0}}D_{t}^{-\alpha }f\left( t\right)$ is the
Riemann-Liouville fractional integral operator which is a direct
extension of Cauchy's multiple integral for arbitrary complex
$\alpha$ with $Re\left( \alpha \right) >0$. A fractional derivative
is then established via a fractional integration and successive
ordinary differential according to
\begin{equation}
_{t_{0}}D_{t}^{\beta }f\left( t\right)
=\frac{d^{n}}{dt^{n}}D_{t}^{\beta -n}f\left( t\right)
\end{equation}
with $Re\left( \beta \right) >0$, and the natural number $n$
satisfies the inequality $n\geq Re\left( \beta \right) >n-1$. Two
special cases are the Riemann-Liouville $_{0}D_{t}^{\beta }$ for $
t_{0}=0$ and Weyl operator $_{-\infty }D_{t}^{\mu }$ for $
t_{0}=-\infty$. The  Riemann-Liouville operator for $ t_{0}=0$ is
given by
\begin{equation}
_{0}D_{t}^{\alpha}f\left( t\right) =\frac{1}{\Gamma \left( n-\alpha\right) }\frac{%
d^{n}}{dt^{n}}\int_{0}^{t}dt^{\prime }\frac{f\left( t^{\prime }\right) }{%
\left( t-t^{\prime }\right) ^{1-n-\alpha}}, \quad n\geq
Re(\alpha)>n-1 .
\end{equation}
For any function $f\left(x,t\right)$ the fractional
Riemann-Liouville differintegration $_{0}D_{t}^{1-\alpha }=\left(
\partial /\partial t\right) _{0}D_{t}^{-\alpha }$ is defined through the
relation
\begin{equation}
_{0}D_{t}^{1-\alpha }f\left( x,t\right) =\frac{1}{\Gamma \left(
\alpha \right) }\frac{\partial }{\partial t}\int_{0}^{t}dt^{\prime
}\frac{f\left( x,t^{\prime }\right) }{\left( t-t^{\prime }\right)
^{1-\alpha }} \ .
\end{equation}
This differintegration operator of an arbitrary power for $ t_{0}=0$
is given by
\begin{equation}
_{0}D_{t}^{\nu }t^{\mu }=\frac{\Gamma \left( 1+\mu \right) }{\Gamma
\left( 1+\mu -\nu \right) }t^{\mu -\nu }
\end{equation}
which coincides with the heuristic generalization of the standard
differentiation
\begin{equation}
\frac{d^{n}t^{m}}{dt^{n}}=\frac{m!}{\left( m-n\right) !}t^{m-n}
\end{equation}
by introduction of Gamma function. An interesting consequence of
Eq.\,(22) is the nonvanishing fractional differintegration of a
constant $a$
\begin{equation}
_{0}D_{t}^{\nu }a=\frac{1}{\Gamma \left( a-\nu \right) }t^{-\nu } \
.
\end{equation}
The Riemann-Liouville differentiation of exponential function leads
to
\begin{equation}
_{0}D_{t}^{\nu }e^{t}=\frac{t^{-\nu }}{\Gamma \left( 1-\nu \right) }%
F_{1,1}\left( 1,1-\nu ,t\right)
\end{equation}
involving the confluent hypergeometric function $F_{1,1}$.


\paragraph{\textbf{Definition 2}} The Caputo fractional derivative
operator \cite{c1Caputo} is given by
\begin{equation}
_{t_{0}}^{C}D_{t}^{\alpha }f\left( t\right) =\frac{1}{\Gamma \left(
\alpha -n\right) }\int_{t_{0}}^{t}dt^{\prime }\frac{f^{\left(
n\right) }\left( t^{\prime }\right) }{\left( t-t^{\prime }\right)
^{\alpha +1-n}}, \quad n-1<\alpha <n
\end{equation}
Under natural condition on the function $f\left( t\right)$, for
$\alpha \rightarrow n$ the Caputo derivative becomes a conventional
$n$-th derivative of the function  $f\left( t\right)$.

The Riemann-Liouville fractional operator (22) and Caputo's
fractional operators (27) have different form. Another difference
between these operators is that the Caputo derivative of a constant
is $0$, whereas the Riemann-Liouville fractional derivative of a
constant is given by Eq.\,(25). On the other hand, the main
advantage of Caputo's approach is that the initial conditions for
fractional differential equations with Caputo derivatives take some
form as for integer-order differential equations.

\paragraph{\textbf{Definition 3}} The Mittag-Leffler function \cite{ml1Mittag,ml2Erdelyi} $E_{\alpha }\left(
z\right)$ is a complex function which depends on on two complex
parameters $\alpha$ and $\beta$. It may be defined by the following
series representation when $\alpha>0$, valid in the whole complex
plane
\begin{equation}
E_{\alpha,\beta}\left( z\right) =\sum_{n=0}^{\infty
}\frac{z^{n}}{\Gamma \left( \alpha n+\beta\right) }, \quad \alpha
>0, \quad z\in C
\end{equation}
In the case $\alpha$ and $\beta$ are real and positive, the series
converges for all values of the argument $z$, so the Mittag-Leffler
function is an entire function. Some special cases of the
Mittag-Leffler function are follow. The Mittag-Leffler function is
the natural generalization of the exponential function. Being a
special case of the Fox function introduced below, it is defined
through the inverse Laplace transform
\begin{equation}
E_{\alpha }\left( -\left( t/\tau \right) ^{\alpha }\right) =\pounds
^{-1}\left\{ \frac{1}{u+\tau ^{-\alpha }u^{1-\alpha }}\right\}
\end{equation}
from which the series expansion
\begin{equation}
E_{\alpha }\left( -\left( t/\tau \right) ^{\alpha }\right)
=\sum_{n=0}^{\infty }\frac{\left( -\left( t/\tau \right) ^{\alpha
}\right) ^{n}}{\Gamma \left( 1+\alpha n\right) }
\end{equation}
can be deduced, where $\tau$ is a constant. The asymptotic behavior
of the Mittag-Leffler function interpolates between the initial
stretched exponential form
\begin{equation}
E_{\alpha }\left( -\left( t/\tau \right) ^{\alpha }\right) \sim \exp
\left(
-\frac{\left( t/\tau \right) ^{\alpha }}{\Gamma \left( 1+\alpha \right) }%
\right)
\end{equation}
for $t< \tau$ and the long-time inverse power-law behavior as
\begin{equation}
E_{\alpha }\left( -\left( t/\tau \right) ^{\alpha }\right) \sim
\left( \frac{t}{\tau }\right) ^{\alpha }\frac{1}{\Gamma \left(
1-\alpha \right) }
\end{equation}
for $t\gg \tau$,  $0<\alpha <1$. Special cases of the Mittag-Leffler
function are the exponential function
\begin{equation}
E_{1}\left( -t/\tau \right) =e^{-t/\tau }
\end{equation}
and the product of the exponential and the complementary error
function
\begin{equation}
E_{1/2}\left( -\left( t/\tau \right) ^{1/2}\right) =e^{-t/\tau }erfc
\left( \left( t/\tau \right) ^{1/2}\right) \ .
\end{equation}

\section{Solution of Gross-Pitaevskii equations of integer and fractional order with HPM}

The principles of HPM and its applicability for various kinds of
differential equations ar given in
\cite{h1Odibat,h2Monami,hpm1He,hpm2He,vol1El,os1He,bif1He,bo1He,o1He,
o2He,o3He,o4He,o5Siddiqui,o6Siddiqui,o7He,o8Abbasbandy,o9Abbasbandy}.
Here after we present a review of the standard HPM and modified HPM
suggested by Momani and Odibat \cite{h1Odibat,h2Monami}. We will
employe HPM to the Gross-Pitaevskii equations of integer order and
fractional order.

\subsection{HPM for Gross-Pitaevskii equation
of integer order}

To explain the basic ideas of this method on nonlinear differential
equation of integer order, we consider one-dimensional
Gross-Pitaevskii equation in the form Eq.\,(1) and adopt the
homotopy perturbation method to this equation. Gross-Pitaevskii
equation (1) is a nonlinear partial differential equation which can
be decomposed as linear and nonlinear part:
\begin{equation}
L\left( \psi \left( x,t\right) \right) +N\left( \psi \left(
x,t\right) \right) =f\left( x,t\right), \quad x\in \Omega, \quad t>0
\end{equation}
where $L( \psi)$ is a linear part and $N( \psi)$ is a nonlinear part
and on the other hand $f\left( x,t\right)$ is known analytical
function of Eq.\,(1). For $f\left( x,t\right)=0$, $L( \psi)$ and
$N(\psi)$ can be defined follow as
\begin{equation}
L\left( \psi \right) =i\hbar \frac{\partial \psi \left( x,t\right) }{%
\partial t},
\quad N\left( \psi \right) =+\frac{\hbar ^{2}}{2m}\nabla ^{2}\psi
\left( x,t\right)+V\left( x\right) \psi \left(
x,t\right)+g\left\vert \psi \left( x,t\right) ^{2}\right\vert \psi
\left( x,t\right) .
\end{equation}
We note that Eq.\,(35) must be defined with the boundary condition
\begin{equation}
B(\psi( x), \ \partial \psi(x)/\partial n)=0,\ x\in \Gamma
\end{equation}
where $\Gamma$ is the boundary of the domain $\Omega$. Homotopy
perturbation method defines the homotopy $\phi \left( x,p\right)
:\Omega \times \left[ 0,1\right] \rightarrow R$ which satisfies
\begin{equation}
H\left( \phi ,p\right) =\left( 1-p\right) \left[ L\left( \phi
\right) -L\left( \psi _{0}\right) \right] +p\left[ L\left( \phi
\right) +N\left( \phi \right) -f\left( x,t\right) \right] =0
\end{equation}
or
\begin{equation}
H\left( \phi ,p\right) =L\left( \phi \right) -L\left( \psi
_{0}\right)
+pL\left( \psi _{0}\right) +p\left[ N\left( \phi \right) -f\left( x,t\right) %
\right] =0
\end{equation}
where $p \in[0,1]$ is an embedding parameter, $\psi_{0}=\psi(x, 0)$
is an initial guess of exact solution $\psi(x,t)$, which is
independent of time and satisfies the boundary conditions. Hence
from Eqs.\,(38) and (39) we obtain
\begin{equation}
H(\phi, 0)=L(\phi)-L(\psi_{0}(x))=0
\end{equation}
\begin{equation}
H(\phi, 1)=L(\psi_{0})+N(\phi)-f(x,t)=0 \ .
\end{equation}
Changing process $p$ from zero to unity is just that of $\phi(x, p)$
from $\psi_{0}(x)$ to $\phi(x,t)$. In topology, this is known as
homotopy \cite{h1Odibat,h2Monami,hpm1He,hpm2He},
$L(\phi)-L(\psi_{0}(x))$ and $L(\psi_{0})+N(\phi)-f(x,t)$ are
homotopic. This basic assumption is that the solution of Eqs.\,(38)
or (39) can be expressed as a power series in $p$:
\begin{equation}
\phi \left( x,p\right) =\phi _{0}\left( x\right) +p\phi _{1}\left(
x\right) +p^{2}\phi _{2}\left( x\right) +...
\end{equation}
The approximation solution of Eqs.\,(1) or (35) can be obtained in
the limit $p\rightarrow 1$ as
\begin{equation}
\psi \left( x,t\right) =\phi \left( x,p\right) _{p\rightarrow
1}=\phi _{0}\left( x\right) +p\phi _{1}\left( x\right) +p^{2}\phi
_{2}\left( x\right) +... \ .
\end{equation}
As a consequence, the final result of standard Gross-Pitaevskii
equation in the form Eq.\,(1) is given
\begin{equation}
\psi \left( x,t\right) =\sum_{j=0}^{\infty }p^{j}\phi _{j}\left(
x\right)
\end{equation}
when $p\rightarrow 1$. Here we note that the solution function $\phi
_{k}\left( x\right)$ includes the constants of the Eq.\,(1). On the
other hand, we say that the convergence of the series (44) has been
proved for many nonlinear ordinary and partial differential
equations \cite{hpm1He,o1He}.

After a brief introduce of HPM, now we can apply HPM to the
Gross-Pitaevskii equation of integer order. To solve Eq.\,(1), using
Eq.\,(38) we construct the following homotopy:
\begin{equation}
\left( 1-p\right) \left( \frac{\partial \phi }{\partial
t}-\frac{\partial
\psi _{0}}{\partial t}\right) +p\left[ \frac{\partial \phi }{\partial t}%
-\frac{i}{\hbar}\left( \frac{\hbar }{2m}\nabla ^{2}\phi -V\left(
x\right) \phi -g\left\vert \phi ^{2}\right\vert \phi \right) \right]
=0 .
\end{equation}
Substituting Eq.\,(42) into Eq.\,(45) and equating the coefficients
of the terms with the identical powers of $p$,
\begin{equation}
p^{0}: \ \frac{\partial \phi _{0}}{\partial t}-\frac{\partial \psi _{0}}{%
\partial t}=0, \quad \phi
_{0}\left( x,0\right)=\psi _{0}\left( x\right)=\psi \left(
x,0\right)
\end{equation}
\begin{equation}
p^{1}: \ \frac{\partial \phi _{1}}{\partial t}-\frac{i}{\hbar}\left(
\frac{\hbar }{2m}\nabla ^{2}\phi _{0}-V\left( x\right) \phi
_{0}-g\left\vert \phi _{0}^{2}\right\vert \phi _{0}\right) =0
\end{equation}
\begin{equation}
p^{2}: \ \frac{\partial \phi _{2}}{\partial t}-\frac{i}{\hbar}\left(
\frac{\hbar }{2m}\nabla ^{2}\phi _{1}-V\left( x\right) \phi
_{1}-g\sum_{i=0}^{1}\sum_{k=0}^{1-i}\left\vert \phi _{i}\right\vert
\left\vert \phi _{k}\right\vert \phi _{1-k-i}\right) =0
\end{equation}
\begin{equation}
p^{3}: \ \frac{\partial \phi _{3}}{\partial t}-\frac{i}{\hbar}\left(
\frac{\hbar }{2m}\nabla ^{2}\phi _{2}-V\left( x\right) \phi
_{2}-g\sum_{i=0}^{2}\sum_{k=0}^{2-i}\left\vert \phi _{i}\right\vert
\left\vert \phi _{k}\right\vert \phi _{1-k-i}\right) =0
\end{equation}
\begin{equation}
\begin{array}{c}
. \\
. \\
.%
\end{array} \nonumber
\end{equation}
\begin{equation}
p^{j}: \ \frac{\partial \phi _{j}}{\partial t}-\frac{i}{\hbar}\left(
\frac{\hbar }{2m}\nabla ^{2}\phi _{j-1}-V\left( x\right) \phi
_{j-1}-g\sum_{i=0}^{j-1}\sum_{k=0}^{j-i-1}\left\vert \phi
_{i}\right\vert \left\vert \phi _{k}\right\vert \phi
_{j-k-i-1}\right) =0
\end{equation}
\begin{equation}
\begin{array}{c}
. \\
. \\
.%
\end{array} \nonumber
\end{equation}
we get the iterative equation
\begin{equation}
\phi _{j}=\frac{i}{\hbar}\int_{0}^{t}\left( \frac{\hbar }{2m}\nabla
^{2}\phi _{j-1}-V\left( x\right) \phi
_{j-1}-g\sum_{i=0}^{j-1}\sum_{k=0}^{j-i-1}\left\vert \phi
_{i}\right\vert \left\vert \phi _{k}\right\vert \phi _{j-k-1}\right)
dt, \quad j=1,2,3,...
\end{equation}
with initial value
\begin{equation}
\phi _{j}\left( x,0\right) =\psi\left( x\right)=\psi\left(
x,0\right).
\end{equation}
Hence all $\phi _{j}$ values can be obtained using Eq.\,(53). The
final solution of Gross-Pitaevskii equation (1) of integer order can
be written in terms of the Eq.\,(53) in the limit $p\rightarrow 1$:
\begin{equation}
\psi \left( x,t\right) =\frac{i}{\hbar}\sum_{j=0}^{\infty }p^{j}\int_{0}^{t}\left( \frac{%
\hbar }{2m}\nabla ^{2}\phi _{j-1}-V\left( x\right) \phi
_{j-1}-g\sum_{i=0}^{j-1}\sum_{k=0}^{j-i-1}\left\vert \phi
_{i}\right\vert \left\vert \phi _{k}\right\vert \phi _{j-k-1}\right)
dt \ .
\end{equation}
HPM supposes that this solution satisfy Eq.\,(1) in the limit of the
$p\rightarrow 1$.

\subsection{HPM for time-fractional Gross-Pitaevskii equation}

To solve the time-fractional Gross-Pitaevskii equation (16), we will
apply HPM to Eq.\,(16). Hence we can construct the following
homotopy
\begin{equation}
\left( 1-p\right) \left( _{0}D_{t}^{\alpha }\phi - \
_{0}D_{t}^{\alpha } \psi _{0}\right) +p\left[ _{0}D_{t}^{\alpha
}\phi -\frac{i}{\hbar }\left( \frac{\hbar }{2m}\nabla ^{2}\phi
-V\left( x\right) \phi -g\left\vert \phi ^{2}\right\vert \phi
\right) \right] =0
\end{equation}
Suppose the solution of Eq.\,(56) to be as following form
\begin{equation}
\phi ^{\alpha }\left( x,p\right) =\phi _{0}^{\alpha }\left( x\right)
+p\phi _{1}^{\alpha }\left( x\right) +p^{2}\phi _{2}^{\alpha }\left(
x\right) +... \ .
\end{equation}
In the limit $p\rightarrow 1$ we can write solution of the
time-fractional equation (16)
\begin{equation}
\psi ^{\alpha }\left( x,t\right) =\sum_{j=0}^{\infty
}p^{j}\phi_{j}^{\alpha}\left( x\right) \ .
\end{equation}
Substituting Eq.\,(57) into Eq.\,(56) and equating the coefficients
of the terms with identical powers of $p$, we obtain following
equations
\begin{equation}
p^{0}: \ _{0}D_{t}^{\alpha }\phi _{0}^{\alpha }-\frac{\partial \psi _{0}^{\alpha }}{%
\partial t}=0
\end{equation}
\begin{equation}
p^{1}: \ _{0}D_{t}^{\alpha }\phi _{1}^{\alpha }-\frac{i}{\hbar }\left( \frac{%
\hbar }{2m}\nabla ^{2}\phi _{0}^{\alpha }-V\left( x\right) \phi
_{0}^{\alpha }-g\left\vert \phi _{0}^{2}\right\vert \phi _{0}\right)
=0
\end{equation}
\begin{equation}
p^{2}: \ _{0}D_{t}^{\alpha }\phi _{2}^{\alpha }-\frac{i}{\hbar }\left( \frac{%
\hbar }{2m}\nabla ^{2}\phi _{1}-V\left( x\right) \phi
_{1}-g\sum_{i=0}^{1}\sum_{k=0}^{1-i}\left\vert \phi _{i}\right\vert
\left\vert \phi _{k}\right\vert \phi _{1-k-i}\right) =0
\end{equation}
\begin{equation}
p^{3}: \ _{0}D_{t}^{\alpha }\phi _{3}^{\alpha }-\frac{i}{\hbar }\left( \frac{%
\hbar }{2m}\nabla ^{2}\phi _{2}-V\left( x\right) \phi
_{2}-g\sum_{i=0}^{2}\sum_{k=0}^{2-i}\left\vert \phi _{i}\right\vert
\left\vert \phi _{k}\right\vert \phi _{1-k-i}\right) =0
\end{equation}
\begin{equation}
\begin{array}{c}
. \\
. \\
.%
\end{array}
\end{equation}
\begin{equation}
p^{j}: \ _{0}D_{t}^{\alpha }\phi _{j}^{\alpha }-i\left( \frac{\hbar }{2m}%
\nabla ^{2}\phi _{j-1}-V\left( x\right) \phi
_{j-1}-g\sum_{i=0}^{j-1}\sum_{k=0}^{j-i-1}\left\vert \phi
_{i}\right\vert \left\vert \phi _{k}\right\vert \phi _{j-k-1}\right)
=0
\end{equation}
\begin{equation}
\begin{array}{c}
. \\
. \\
.%
\end{array}
\end{equation}
Finally we can write iterative relation as
\begin{equation}
\phi _{j}^{\alpha }=\frac{i}{\hbar \Gamma \left( \alpha \right) }%
\int_{0}^{t}\left( t-\tau \right) ^{\alpha -1}\left( \frac{\hbar
}{2m}\nabla ^{2}\phi _{j-1}-V\left( x\right) \phi
_{j-1}-g\sum_{i=0}^{j-1}\sum_{k=0}^{j-i-1}\left\vert \phi
_{i}\right\vert \left\vert \phi _{k}\right\vert \phi _{j-k-1}\right)
d\tau
\end{equation}
with initial value
\begin{equation}
\phi _{j}^{\alpha }\left( x,0\right) =\psi\left( x\right)=\psi\left(
x,0\right), \quad j=1,2,3,... \ .
\end{equation}
The final solution of time-fractional Gross-Pitaevskii equation (16)
can be given in terms of the Eq.\,(66) in the limit $p\rightarrow
1$:
\begin{equation}
\psi^{\alpha } \left( x,t\right) =\frac{i}{\hbar \Gamma \left( \alpha \right) }%
\sum_{j=0}^{\infty }p^{j}\int_{0}^{t}\left( t-\tau \right) ^{\alpha
-1}\left( \frac{\hbar }{2m}\nabla ^{2}\phi _{j-1}^{\alpha}-V\left(
x\right) \phi
_{j-1}^{\alpha}-g\sum_{i=0}^{j-1}\sum_{k=0}^{j-i-1}\left\vert \phi
_{i}^{\alpha}\right\vert \left\vert \phi_{k}^{\alpha}\right\vert
\phi_{j-k-1}^{\alpha }\right) d\tau
\end{equation}
HPM supposes that this solution satisfy the Eq.\,(16) in the limit
of the $p\rightarrow 1$.

\section{Numerical implementations for different potentials}

Now we will investigate GP end time-fractional GP equations to
discuss the ground state time-dependent dynamic of the Bose-Einstein
condensation of weakly interacting system which involve attracting
and repulsive interactions for different potential. In analysis we
will use HPM to solve GP equations.


\paragraph{\textbf{Example 1}} Firstly we consider GP equation of integer and fractional order
for $V\left( x\right) =0$ and repulsive interaction $g=1>0$. Hence
we will discuss GPE of integer order in \emph{1.Case-I} and GPE of
fractional order in \emph{1.Case-II}.

\paragraph{1.Case-I} For $V\left( x\right) =0$ and repulsive interaction $g=1>0$,
the iterative equation (53) is given as
\begin{equation}
\phi _{j}=\frac{i}{\hbar }\int_{0}^{t}\left( \frac{\hbar }{2m}\nabla
^{2}\phi _{j-1}-g\sum_{i=0}^{j-1}\sum_{k=0}^{j-i-1}\left\vert \phi
_{i}\right\vert \left\vert \phi _{k}\right\vert \phi _{j-k-1}\right)
dt \ .
\end{equation}
For $g=1>0$, the HPM solutions of $\phi _{j}$ are given by
\begin{equation}
\phi _{0}=\tanh x
\end{equation}
\begin{equation}
\phi _{1}=-\frac{-it}{\Gamma \left( 2\right)}\tanh x
\end{equation}
\begin{equation}
\phi _{2}=-\frac{t^{2}}{\Gamma \left( 3\right)}\tanh x
\end{equation}
\begin{equation}
\phi _{3}=-\frac{it^{3}}{\Gamma \left( 4\right)}\tanh x
\end{equation}
\begin{equation}
\begin{array}{c}
. \\
. \\
.%
\end{array}
\end{equation}
\begin{equation}
\phi _{n}=\left( -1\right) ^{n}\frac{\left( it\right) ^{n}}{\Gamma
\left( n+1\right)}\tanh x
\end{equation}
where we set $\hbar=m=1$ for simplicity. If we put these terms in
the series expansion
\begin{equation}
\psi\left( x,t\right) =\phi _{0}\left( x\right) +p\phi _{1}\left(
x\right) +p^{2}\phi _{2}\left( x\right) +...=\sum_{j=0}^{\infty
}p^{j}\phi _{j}\left( x\right)
\end{equation}
we obtain
\begin{equation}
\psi _{g>0}\left( x,t\right) =\tanh x + \sum_{n=1}^{\infty }\left(
-1\right) ^{n}\frac{\left( it\right) ^{n}}{\Gamma \left(
n+1\right)}\tanh x \ .
\end{equation}
This solution can be represented in terms of Mittag-Leffler function
for $\alpha=1$ as
\begin{equation}
\psi _{g>0}\left( x,t\right) =E_{1}\left[ -\left( \frac{t}{\tau }\right) %
\right] \tanh x=\tanh xe^{-t/\tau }
\end{equation}
where $i=1/\tau$. This unnormalized solution is of the same
analytical form with Eq.\,(11) which corresponds \emph{dark soliton}
solution of GP equation. The numerical demonstration of the
Eq.\,(78) fot $t=0$ and the probability density of $\psi
_{g>0}\left( x,t\right)$ are given is given in Figs.\,(1a) and (1b),
respectively. As it can be seen from Fig.\,(1b) when external
potential  $V\left( x\right) =0$, the probability density for
attracting interactions has \emph{dark soliton} form. These results
confirm that obtained solution using HPM is the same with analytical
solution of GP equation of integer order for $V\left( x\right) =0$
and $g>0$.
\begin{figure*}[ht] \label{Figure 1}
\begin{center}
\includegraphics[width=5.5 in]{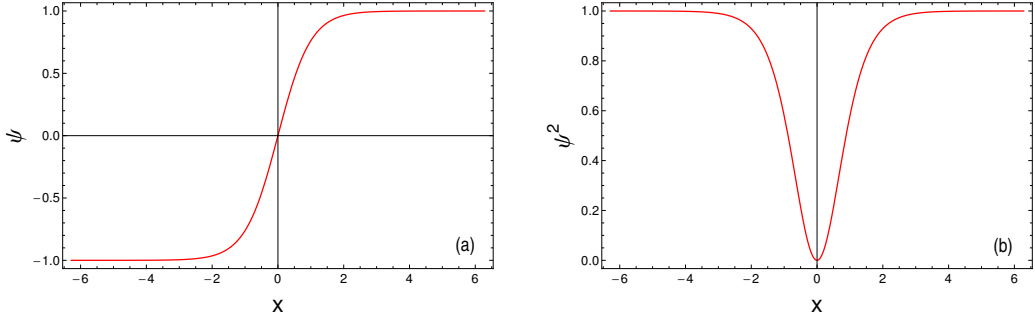}
\caption{In the case $V\left(x\right)=0$ and $g>0$ a) Numerical
representation of the $\psi_{g>0}\left(x,0\right)$, b) Numerical
representation of the probability density of
$\psi_{g>0}\left(x,t\right)$.}
\end{center}
\end{figure*}
\paragraph{1.Case-II} Now we consider time-fractional GP equation for $V\left( x\right)
=0$ and repulsive interaction i.e. $g>0$. The iterative equation
(66) of the time-fractional GP equation for $V\left(x\right)=0$ and
$g=1>0$ is given by
\begin{equation}
\phi _{j}^{\alpha }=\frac{i}{\hbar \Gamma \left( \alpha \right) }%
\int_{0}^{t}\left( t-\tau \right) ^{\alpha -1}\left( \frac{\hbar
}{2m}\nabla ^{2}\phi _{j-1}^{\alpha
}-g\sum_{i=0}^{j-1}\sum_{k=0}^{j-i-1}\left\vert \phi _{i}^{\alpha
}\right\vert \left\vert \phi _{k}^{\alpha }\right\vert \phi
_{j-k-1}^{\alpha }\right) d\tau \ .
\end{equation}
In this case, the HPM solutions of $\phi _{j}^{\alpha }$ are given
by
\begin{equation}
\phi _{0}^{\alpha }=\tanh x
\end{equation}
\begin{equation}
\phi _{1}^{\alpha }=-\frac{it^{\alpha }}{\Gamma \left( \alpha +1\right) }%
\tanh x
\end{equation}
\begin{equation}
\phi _{2}^{\alpha }=-\frac{t^{2\alpha }}{\Gamma \left( 2\alpha +1\right) }%
\tanh x
\end{equation}
\begin{equation}
\phi _{3}^{\alpha }=\frac{it^{3\alpha }}{\Gamma \left( 3\alpha +1\right) }%
\tanh x
\end{equation}
\begin{equation}
\begin{array}{c}
. \\
. \\
.%
\end{array}
\end{equation}
\begin{equation}
\phi _{n}^{\alpha }=\left( -1\right) ^{n}\frac{\left( it^{\alpha
}\right) ^{n}}{\Gamma \left( n\alpha +1\right) }\tanh x
\end{equation}
where we set $\hbar=m=1$ for simplicity. If we put these terms in
the series expansion
\begin{equation}
\psi^{\alpha }\left( x,t\right) =\phi _{0}^{\alpha }\left( x\right)
+p\phi _{1}^{\alpha }\left( x\right) +p^{2}\phi _{2}^{\alpha }\left(
x\right) +...=\sum_{j=0}^{\infty }p^{j}\phi _{j}^{\alpha }\left(
x\right)
\end{equation}
we obtain final result
\begin{equation}
\psi^{\alpha}_{g>0}\left( x,t\right) =\tanh x+\sum_{n=1}^{\infty
}\left( -1\right)
^{n}\frac{\left( it^{\alpha }\right) ^{n}}{\Gamma \left( n\alpha +1\right) }%
\tanh x \ .
\end{equation}
This solution can be written in terms of Mittag-Leffler function for
arbitrary $\alpha$ value as
\begin{equation}
\psi^{\alpha}_{g>0}\left( x,t\right) =\tanh x+\sum_{n=1}^{\infty
}\frac{\left( -\left( t/\tau \right) ^{\alpha }\right) ^{n}}{\Gamma
\left( n \alpha+1 \right) }\tanh x=E_{\alpha }\left( -\left( t/\tau
\right) ^{\alpha }\right)\tanh x
\end{equation}
where we set  $i=1/\tau^\alpha$. According to Eqs.\,(31) and (32)
the asymptotic behavior of the Mittag-Leffler function, the
fractional solution (88) is given by stretched exponential form
\begin{equation}
\psi^{\alpha}_{g>0}\left( x,t\right) =E_{\alpha }\left( -\left(
t/\tau \right) ^{\alpha }\right)\tanh x  \sim  \exp \left(
-\frac{\left( t/\tau \right) ^{\alpha }}{\Gamma \left( \alpha+1
\right) }\right)\tanh x
\end{equation}
or for long-time regime is given by inverse power-law as
\begin{equation}
\psi^{\alpha}_{g>0}\left( x,t\right) = E_{\alpha }\left( -\left(
t/\tau \right) ^{\alpha }\right)\tanh x \sim  \left( \frac{t}{\tau
}\right) ^{\alpha }\frac{1}{\Gamma \left( 1-\alpha \right) } \tanh x
\ .
\end{equation}
The nice analytical results are solutions of the time-fractional GP
equation for $V\left(x\right)=0$ and $g=1>0$. It can be clearly seen
from Eqs.\,(89) and (90) that the time-dependent solution of
time-fractional GP equation (16) is different from standard GP
equation (1). Indeed, solutions (89) and (90) of time-fractional GP
equation indicate that the ground state dynamics of the
Bose-Einstein condensation evolute with time in complex space as
obey to stretched exponential for short time regime and power law
for long-time regime in the case external potential zero and
interactions between bosonic particles are repulsive. Whereas for
the same case i.e. $V\left(x\right)=0$ and $g=1>0$ the ground state
dynamics of the condensation exponentially evolutes with time.
However, the spatial part of both solution in Eq.\,(88) and in
Eq.\,(78) are equal. Therefore, it is excepted that the
time-fractional dynamics of condensation also produces \emph{dark
soliton} behavior similar to Eq.\,(78) as it can be seen in
Fig.\,(1b). On the other hand, here we must remark that the
fractional parameter $\alpha$ is a measure of fractality in between
particles in condensation progress. Hence it determine time
evolution of the ground state dynamics of condensation.

In order to demonstrate the effect of the fractional parameter on
the ground state dynamics or another say to investigate the ground
state dynamics of the fractional condensation we plot the real and
imaginer part of the $\psi^{\alpha}_{g>0}\left( x,t\right)$ for
several $\alpha$ values.
\begin{figure*}[ht] \label{Figure 2}
\begin{center}
\subfigure[\hspace{-1.3cm}]{\label{--}
\includegraphics[width=2.9 in]{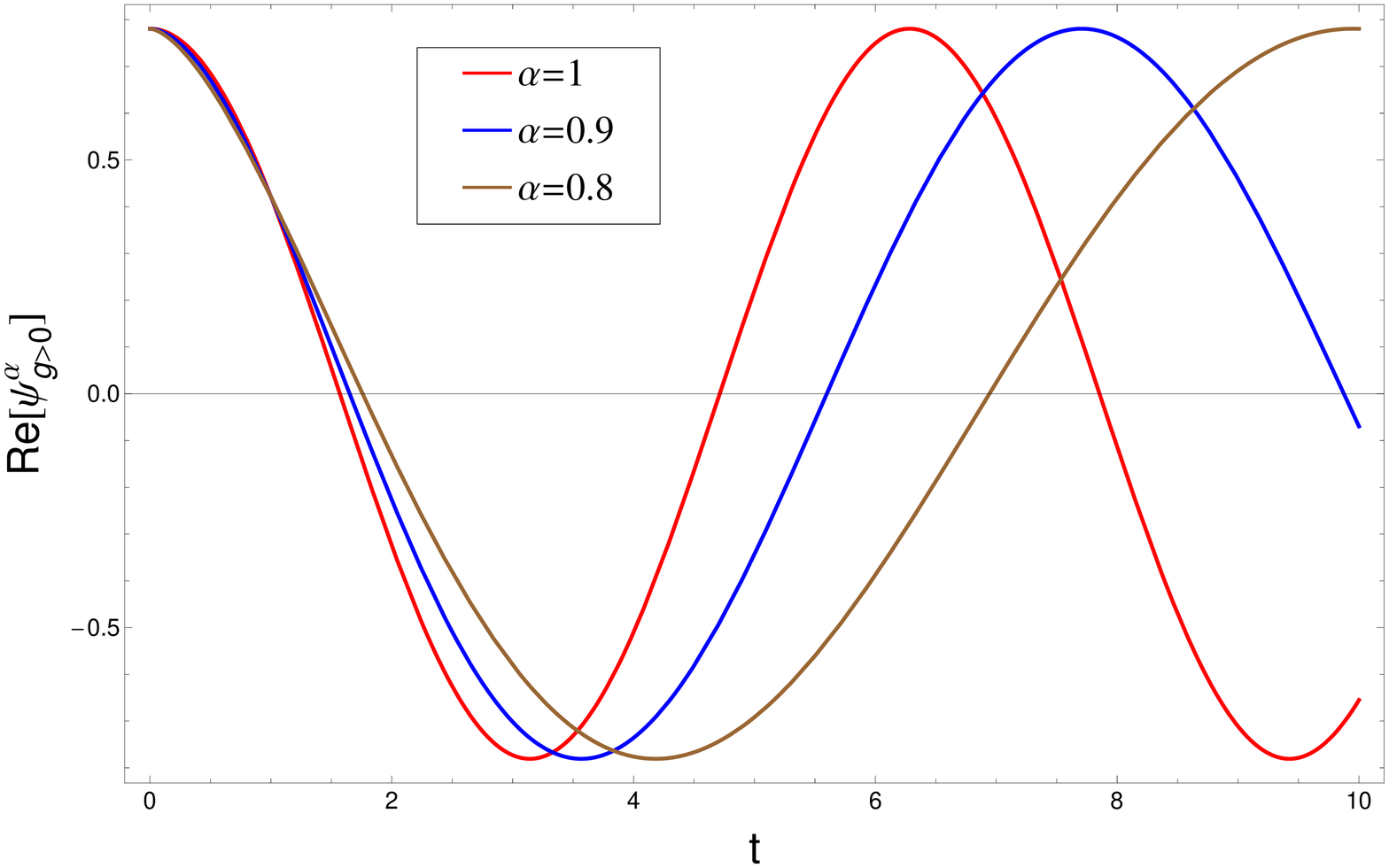}}
\subfigure[\hspace{-1.3cm}]{\label{--}
\includegraphics[width=2.9 in]{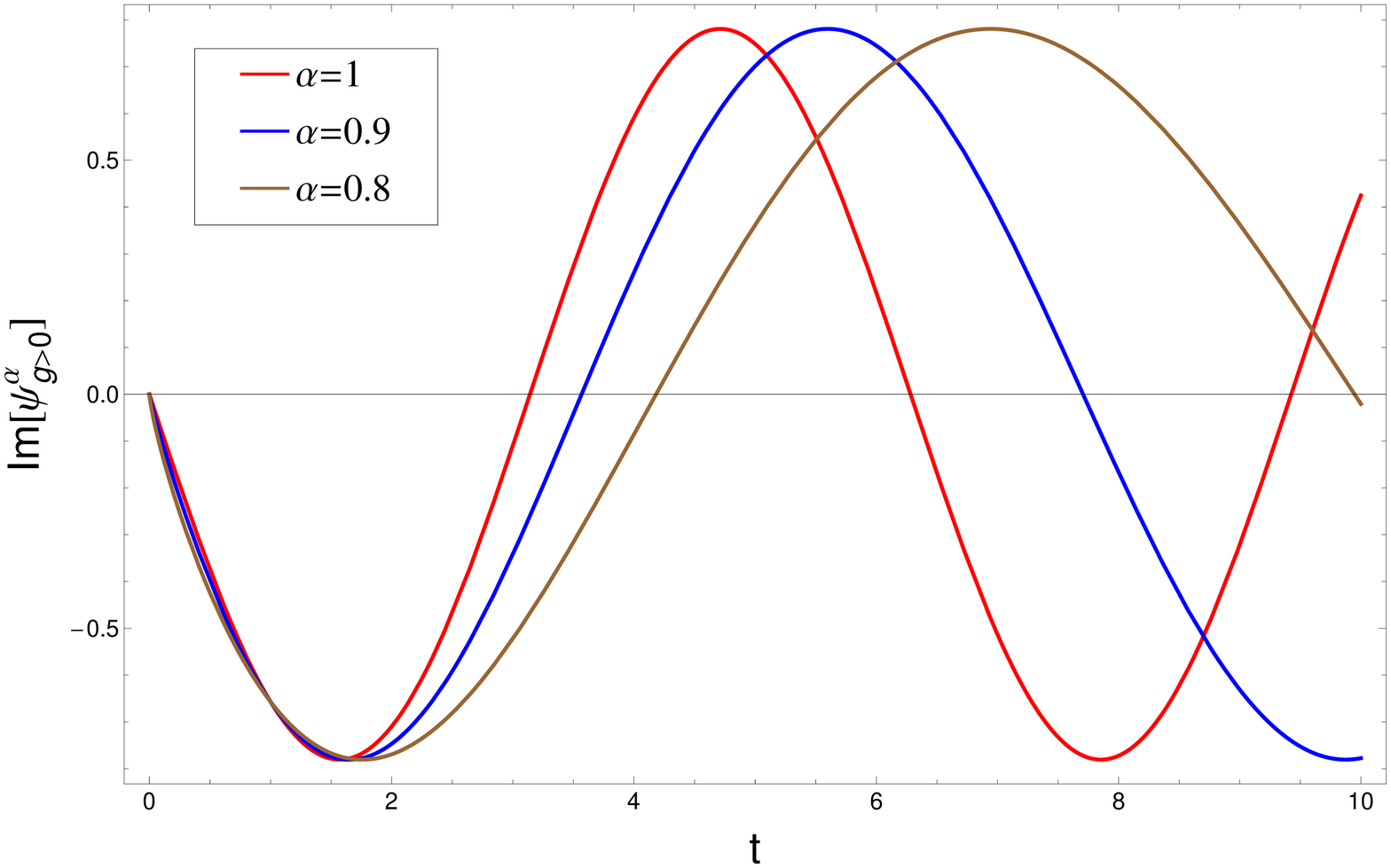}}
\caption{In the case $V\left(x\right)=0$ and $g>0$ a) Real part of
the $\psi^{\alpha}_{g>0}\left( x,t\right)$ for $\alpha=1, 0.9, 0.8$
values, b) Imaginer part of the $\psi^{\alpha}_{g>0}\left(
x,t\right)$ for $\alpha=1, 0.9, 0.8$ values.}
\end{center}
\end{figure*}
As it can be seen from Figure 2 that the time evolution of the real
and imaginer part of the wave function $\psi^{\alpha}_{g>0}\left(
x,t\right)$ clearly depend on the fractional parameter $\alpha$. For
small values of the time, wave function solution depend on time
nearly coincident for different $\alpha$ values, however fractional
parameter $\alpha$ values substantially affect the solution of
$\psi^{\alpha}_{g>0}\left( x,t\right)$ when time is increased.
Indeed for $\alpha<1$ values this effect can be clearly seen in
Figure 2 (a) and (b) at large time values.
\begin{figure*}\label{Figure 3}
\centering{
\includegraphics[width=3.5in]{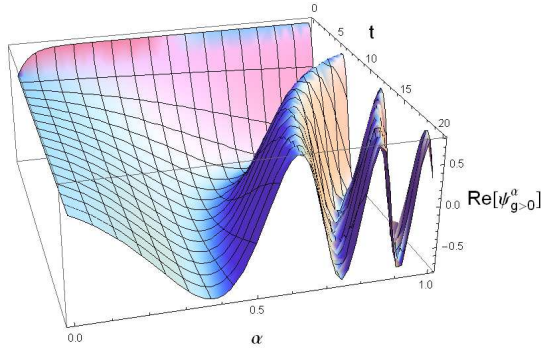}
\caption{Time and fractional parameter $\alpha$  dependence of the
real part of $\psi^{\alpha}_{g>0}\left( x,t\right)$ for
$V\left(x\right)=0$ and $g>0$.}}
\end{figure*}

On the other hand to see effect of the fractional parameter on the
ground state dynamics of condensation of bosonic particles with
fractality, we give the surface plot in Figure 3. This figure
clearly shows that the real part of the wave function
$\psi^{\alpha}_{g>0}\left( x,t\right)$ evolutes to make oscillation
with time for arbitrary $\alpha$ values, however, for in the limit
$\alpha \rightarrow 0$, at the same time, or in the limit $t
\rightarrow 0$ the wave function $\psi^{\alpha}_{g>0}\left(
x,t\right)$ goes to stationary state which means that the wave
function in these limit does not change. Hence in these limit
values, stationary solution of the time-fractional GP equation is
independent of time. For this reason, the soliton solution of time
fractional GP equation (16) in these limit is stable similar to
soliton solution of GP equation (1). The same behavior appears in
imaginer part of the the wave function $\psi^{\alpha}_{g>0}\left(
x,t\right)$ .



\paragraph{\textbf{Example 2}} Secondly we consider GP equation of integer and fractional order for
$V\left( x\right) =0$ and attractive interaction $g=-1<0$. Hence we
will discuss GPE of integer order in \emph{2.Case-I} and GPE of
fractional order in \emph{2.Case-II}.
\paragraph{2.Case-I} For $V\left( x\right) =0$ and attractive interaction $g=-1<0$,
the iterative equation (53) is given as
\begin{equation}
\phi _{j}=\frac{i}{\hbar }\int_{0}^{t}\left( \frac{\hbar }{2m}\nabla
^{2}\phi _{j-1}-g\sum_{i=0}^{j-1}\sum_{k=0}^{j-i-1}\left\vert \phi
_{i}\right\vert \left\vert \phi _{k}\right\vert \phi _{j-k-1}\right)
dt \ .
\end{equation}
In this case, the HPM solutions of $\phi _{j}$ are given by
\begin{equation}
\phi _{0}=\sec hx
\end{equation}
\begin{equation}
\phi _{1}=-\frac{it}{2\Gamma \left( 2\right)}\sec hx
\end{equation}
\begin{equation}
\phi _{2}=-\frac{t^{2}}{3\Gamma \left( 3\right)}\sec hx
\end{equation}
\begin{equation}
\phi _{3}=\frac{it^{3}}{4\Gamma \left( 4\right)}\sec hx
\end{equation}
\begin{equation}
\begin{array}{c}
. \\
. \\
.%
\end{array}
\end{equation}
\begin{equation}
\phi _{n}=\left( -1\right) ^{n}\frac{\left( it\right)
^{n}}{(n+1)\Gamma \left( n+1\right)}\sec hx
\end{equation}
where we set $\hbar=m=1$ for simplicity. Using the series expansion
\begin{equation}
\psi \left( x,t\right) =\phi _{0}\left( x\right) +p\phi _{1}\left(
x\right) +p^{2}\phi _{2}\left( x\right) +...=\sum_{j=0}^{\infty
}p^{j}\phi _{j}\left( x\right)
\end{equation}
the final results can be written as
\begin{equation}
\psi _{g<0}\left( x,t\right) =\sec hx+\sum_{n=1}^{\infty }\left(
-1\right) ^{n}\frac{\left( it\right) ^{n}}{(n+1)\Gamma \left(
n+1\right)}\sec hx \ .
\end{equation}
This solution can be defined Mittag-Leffler function for $\alpha=1$
as
\begin{equation}
\psi _{g<0}\left( x,t\right) \cong E_{1}\left[ -\left( \frac{t}{\tau }\right) %
\right] \sec hx=\sec hxe^{-t/\tau }
\end{equation}
where $i=1/\tau$. This unnormalized solution is of the same
analytical form with Eq.\,(13) which corresponds \emph{bright
soliton} solution of GP equation. The numerical demonstration of the
Eq.\,(100) for $t=0$ and the probability density of $\psi
_{g<0}\left( x,t\right)$ are given is given in Figs.\,(4a) and (4b),
respectively. As it can be seen from Fig.\,(4b) when external
potential  $V\left( x\right) =0$, the probability density for
attractive interactions has \emph{bright soliton} form. These
results confirm that obtained solution using HPM is the same with
analytical solution of GP equation of integer order for $V\left(
x\right) =0$ and $g<0$.
\begin{figure*}[ht] \label{Figure 4}
\begin{center}
\includegraphics[width=5.5 in]{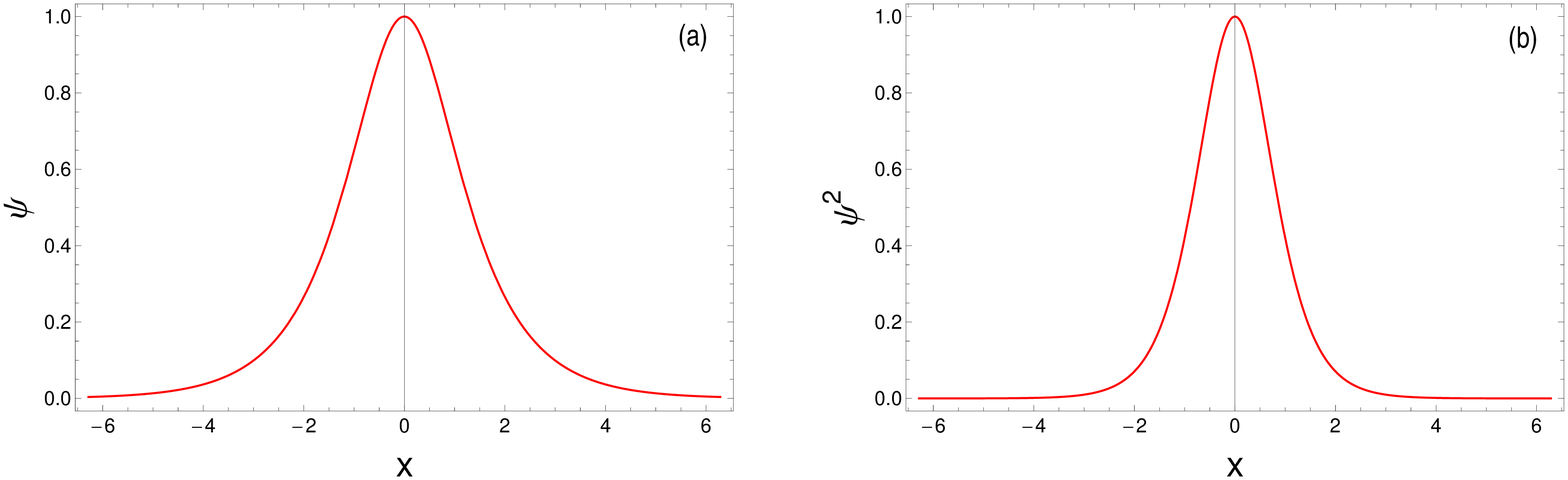}
\caption{In the case $V\left(x\right)=0$ and $g<0$ a) Numerical
representation of the $\psi_{g<0}\left(x,0\right)$, b) Numerical
representation of the probability density of
$\psi_{g<0}\left(x,t\right)$.}
\end{center}
\end{figure*}

\paragraph{2.Case-II}
Now we consider time-fractional GP equation for $V\left( x\right)
=0$ and $g=-1<0$, The iterative equation (66) of the time-fractional
GP equation for $V\left(x\right)=0$
\begin{equation}
\phi _{j}^{\alpha }=\frac{i}{\hbar \Gamma \left( \alpha \right) }%
\int_{0}^{t}\left( t-\tau \right) ^{\alpha -1}\left( \frac{\hbar
}{2m}\nabla ^{2}\phi _{j-1}^{\alpha
}-g\sum_{i=0}^{j-1}\sum_{k=0}^{j-i-1}\left\vert \phi _{i}^{\alpha
}\right\vert \left\vert \phi _{k}^{\alpha }\right\vert \phi
_{j-k-1}^{\alpha }\right) d\tau \ .
\end{equation}
For attractive interaction, the HPM solutions of $\phi _{j}^{\alpha
}$ are given by
\begin{equation}
\phi _{0}^{\alpha }=\sec hx
\end{equation}
\begin{equation}
\phi _{1}^{\alpha }=-\frac{it^{\alpha }}{\Gamma \left( \alpha +2\right) }%
\sec hx
\end{equation}
\begin{equation}
\phi _{2}^{\alpha }=-\frac{t^{2\alpha }}{\Gamma \left( 2\alpha +2\right) }%
\sec hx
\end{equation}
\begin{equation}
\phi _{3}^{\alpha }=\frac{it^{3\alpha }}{\Gamma \left( 3\alpha +2\right) }%
\sec hx
\end{equation}
\begin{equation}
\begin{array}{c}
. \\
. \\
.%
\end{array}
\end{equation}
\begin{equation}
\phi _{n}^{\alpha }=\left( -1\right) ^{n}\frac{\left( it^{\alpha
}\right) ^{n}}{\Gamma \left( n\alpha +2\right) }\sec hx
\end{equation}
where we set $\hbar=m=1$ for simplicity. All solutions can be added
in series expansion as
\begin{equation}
\psi ^{\alpha }\left( x,t\right) =\phi _{0}^{\alpha }\left( x\right)
+p\phi _{1}^{\alpha }\left( x\right) +p^{2}\phi _{2}^{\alpha }\left(
x\right) +...=\sum_{j=0}^{\infty }p^{j}\phi _{j}^{\alpha }\left(
x\right) \ .
\end{equation}
This produces the final results for $g<0$
\begin{equation}
\psi_{g<0}^{\alpha }\left( x,t\right) =\sec hx+\sum_{n=1}^{\infty
}\left( -1\right)
^{n}\frac{\left( it^{\alpha }\right) ^{n}}{\Gamma \left( n\alpha +2\right) }%
\sec hx \ .
\end{equation}
This solution can be written in terms of Mittag-Leffler function as
\begin{equation}
\psi _{g<0}^{\alpha }\left( x,t\right) =\sec hx +\sum_{n=1}^{\infty
}\frac{\left( -\left(
t/\tau \right) ^{\alpha }\right) ^{n}}{(n\alpha+1)\Gamma \left( n\alpha+1 \right) }\sec hx%
\cong E_{\alpha }\left( -\left( t/\tau \right) ^{\alpha }\right)\sec
hx
\end{equation}
where we set  $i=1/\tau^\alpha$. According to Eqs.\,(31) and (32)
the asymptotic behavior of the Mittag-Leffler function, the
fractional solution (110) is given by stretched exponential form
\begin{equation}
\psi _{g<0}^{\alpha }\left( x,t\right) \cong E_{\alpha }\left(
-\left( t/\tau \right) ^{\alpha }\right)\sec hx \sim \exp \left(
-\frac{\left( t/\tau \right) ^{\alpha }}{\Gamma \left( \alpha+1
\right) }\right)\sec hx
\end{equation}
or for long-time regime is given by inverse power-law as
\begin{equation}
\psi _{g<0}^{\alpha }\left( x,t\right) \cong E_{\alpha }\left(
-\left( t/\tau \right) ^{\alpha }\right)\sec hx \sim \left(
\frac{t}{\tau }\right) ^{\alpha }\frac{1}{\Gamma \left( 1-\alpha
\right) }\sec hx \ .
\end{equation}
The nice analytical results are solutions of the time-fractional GP
equation for $V\left(x\right)=0$ and $g=-1<0$. It can be clearly
seen from Eqs.\,(111) and (112) that the time-dependent solution of
time-fractional GP equation (16) is different from standard GP
equation (1). Indeed, solutions (111) and (112) of time-fractional
GP equation indicate that the ground state dynamics of the
Bose-Einstein condensation evolute with time in complex space as
obey to stretched exponential for short time regime and power law
for long-time regime in the case external potential zero and
interactions between bosonic particles are attractive. Whereas for
the same case i.e. $V\left(x\right)=0$ and $g=-1<0$ the ground state
dynamics of the condensation exponentially evolutes with time.
However, the spatial part of both solution in Eq.\,(110) and in
Eq.\,(100) are equal. Therefore, it is excepted that the
time-fractional dynamics of condensation also produces \emph{bright
soliton} behavior similar to Eq.\,(100) as it can be seen in
Fig.\,(4b). On the other hand, here we must remark that the
fractional parameter $\alpha$ is a measure of fractality in between
particles in condensation progress. Hence it determine time
evolution of the ground state dynamics of condensation.
\begin{figure*}[ht] \label{Figure 5}
\begin{center}
\subfigure[\hspace{-1.1cm}]{\label{--}
\includegraphics[width=2.9 in]{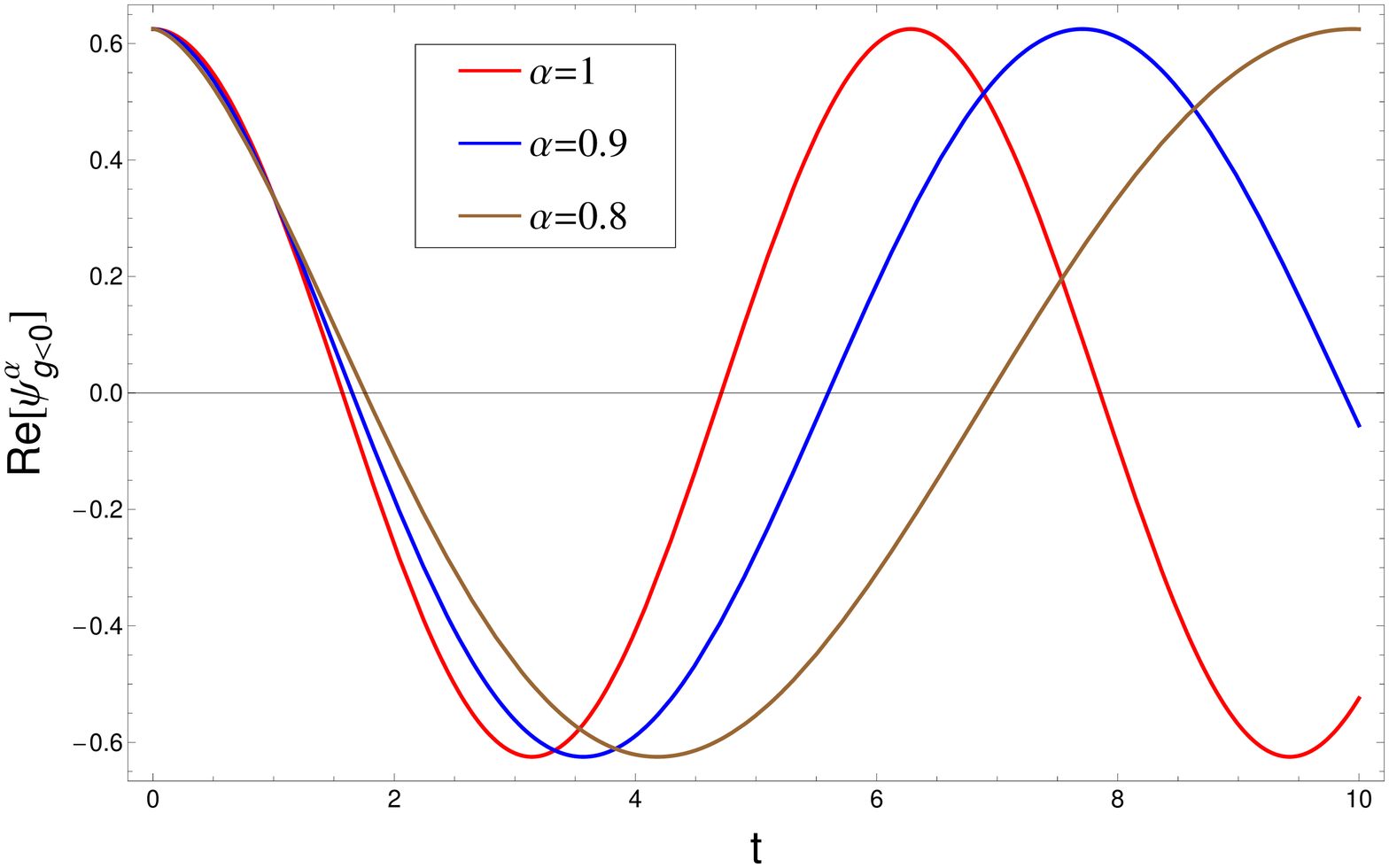}}
\hspace{0.01cm} \subfigure[\hspace{-1.1cm}]{\label{--}
\includegraphics[width=2.9 in]{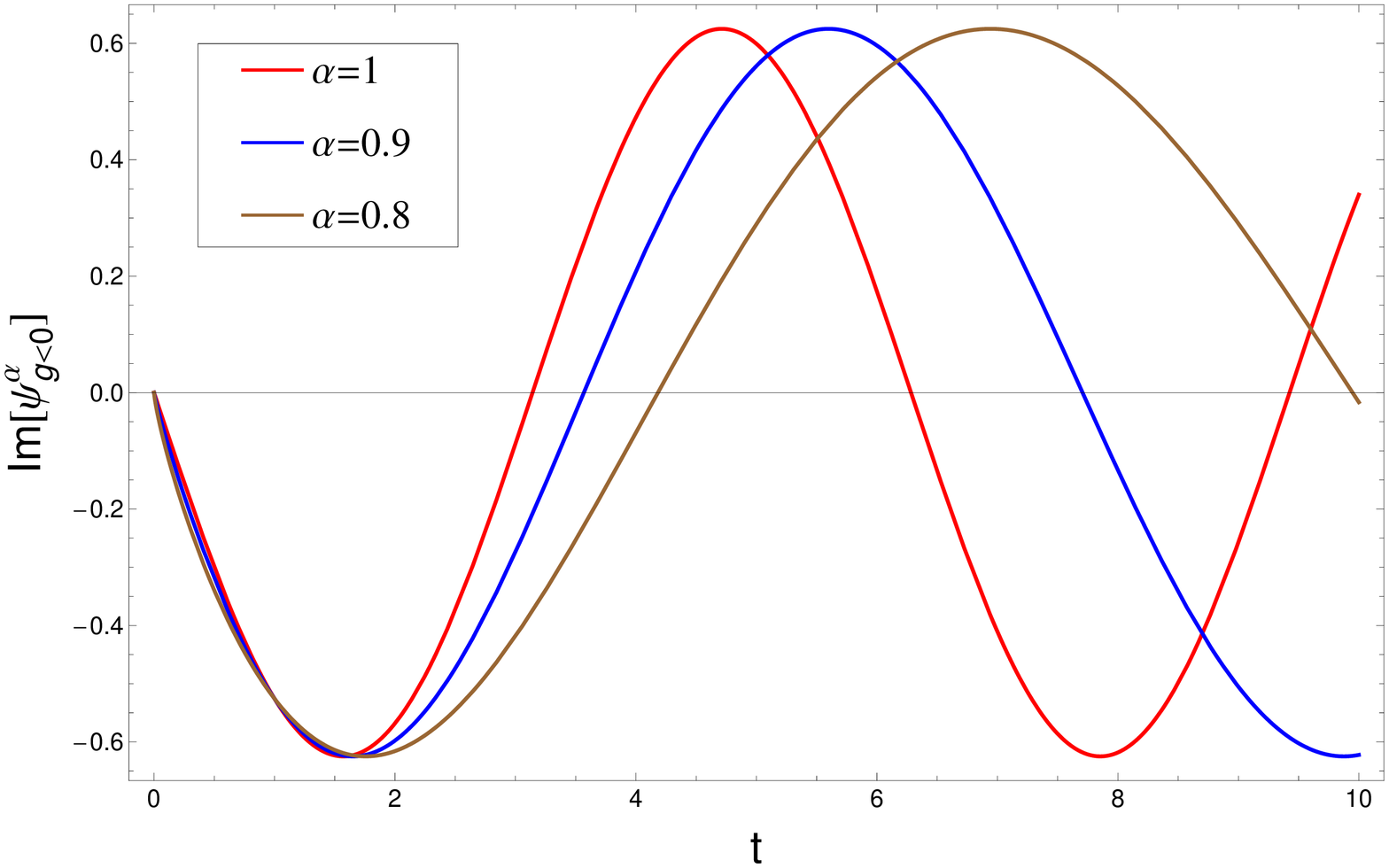}}
\hspace{0.01cm} \caption{In the case $V\left(x\right)=0$ and $g<0$
a) Real part of the $\psi^{\alpha}_{g<0}\left( x,t\right)$ for
$\alpha=1, 0.9, 0.8$ values, b) Imaginer part of the
$\psi^{\alpha}_{g<0}\left( x,t\right)$ for $\alpha=1, 0.9, 0.8$
values.}
\end{center}
\end{figure*}

In order to demonstrate the effect of the fractional parameter on
the ground state dynamics or another say to investigate ground state
dynamics of fractional condensation we plot the real and imaginer
part of the $\psi^{\alpha}_{g<0}\left( x,t\right)$ for several
$\alpha$ values.
\begin{figure*}\label{Figure 6}
\centering{
\includegraphics[width=3.5in]{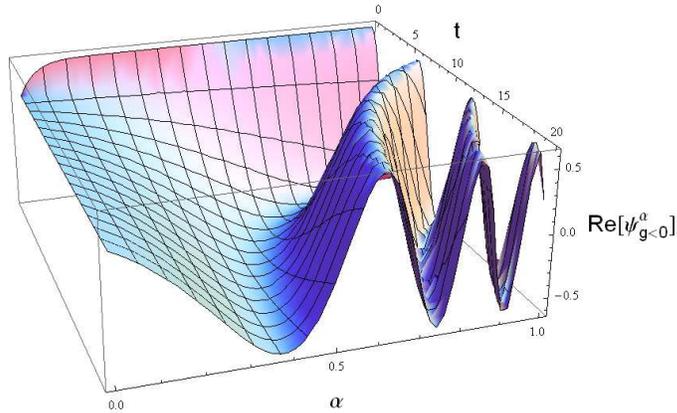}
\caption{Time and fractional parameter $\alpha$  dependence of the
real part of $\psi^{\alpha}_{g<0}\left(x,t\right)$ for
$V\left(x\right)=0$ and $g<0$.}}
\end{figure*}
As it can be seen from Figure 5 that the time evolution of the real
and imaginer part of the wave function $\psi^{\alpha}_{g<0}\left(
x,t\right)$ clearly depend on fractional parameter $\alpha$. For
small values of the time, wave function solution depend on time
nearly coincident for different $\alpha$ values, however fractional
parameter $\alpha$ values substantially affect the solution of
$\psi^{\alpha}_{g<0}\left( x,t\right)$ when time is increased.
Indeed for $\alpha<1$ values this effect can be clearly seen in
Figure 5 (a) and (b) at large time values. Furthermore, to see
effect of the fractional parameter on the ground state dynamics of
condensation of bosonic particles, we give the surface plot in
Figure 6. This figure clearly shows that the real part of the wave
function $\psi^{\alpha}_{g<0}\left(x,t\right)$ evolutes to make
oscillation with time for arbitrary $\alpha$ values, however, for in
the limit $\alpha \rightarrow 0$, at the same time, or in the limit
$t \rightarrow 0$ the wave function $\psi^{\alpha}_{g<0}\left(
x,t\right)$ goes to stationary state which means that the wave
function in these limit does not change. Hence in these limit
values, stationary solution of the time-fractional GP equation is
independent of time. For this reason, the soliton solution of time
fractional GP equation (16) in these limit is stable similar to
soliton solution of GP equation (1). The same behavior appears in
imaginer part of the the wave function $\psi^{\alpha}_{g<0}\left(
x,t\right)$.


\paragraph{\textbf{Example 3}} Thirdly, in this example, by using HPM we will
obtain solutions of the GP equation of integer and fractional order
for optical lattice potential $V\left( x\right) =\sin ^{2}x$ and
repulsive interaction i.e. $g=1>0$. Hence we will consider GPE of
integer order in \emph{3.Case-I} and GPE of fractional order in
\emph{3.Case-II}. We note that we have found the analytical solution
of GP equation as $\psi _{g=1}\left( x,t\right) =e^{-3it/2}\cos x$
in the case $g=1>0$ for optical lattice potential $V\left( x\right)
=\sin ^{2}x$ which satisfy Eq.\,(1).

\paragraph{3.Case-I} For optical lattice potential $V\left( x\right) =\sin ^{2}x$ and
repulsive interaction $g=1>0$, the iterative equation (53) is given
as
\begin{equation}
\phi _{j}=\frac{i}{\hbar }%
\int_{0}^{t}\left( \frac{\hbar }{2m}\nabla ^{2}\phi _{j-1}-\sin
^{2}x\phi _{j-1}-g\sum_{i=0}^{j-1}\sum_{k=0}^{j-i-1}\left\vert \phi
_{i}\right\vert \left\vert \phi _{k}\right\vert \phi _{j-k-1}\right)
d\tau
\end{equation}
For this potential, the HPM solutions of $\phi _{j}$ are given by
\begin{equation}
\phi _{0}=\cos x
\end{equation}
\begin{equation}
\phi _{1}=-\frac{3it}{2\Gamma \left(2\right) }%
\cos x
\end{equation}
\begin{equation}
\phi _{2}=-\frac{9t^2}{4\Gamma \left( 3\right) }%
\cos x
\end{equation}
\begin{equation}
\phi _{3}=\frac{27it^{3}}{8\Gamma \left( 4\right) }%
\cos x
\end{equation}
\begin{equation}
\begin{array}{c}
. \\
. \\
.%
\end{array}
\end{equation}
\begin{equation}
\phi _{n}=\left( -1\right) ^{n}\frac{\left( 3it\right)
^{n}}{2^n\Gamma \left( n+1\right) }\cos x
\end{equation}
where we set $\hbar=m=1$ for simplicity. If these $\phi _{j}$ terms
can be put in the series
\begin{equation}
\psi\left( x,t\right) =\phi _{0}\left( x\right) +p\phi _{1}\left(
x\right) +p^{2}\phi _{2}\left( x\right) +...=\sum_{j=0}^{\infty
}p^{j}\phi _{j}\left( x\right)
\end{equation}
we obtain final results for $g=1>0$
\begin{equation}
\psi _{g>0}\left( x,t\right) =\cos x+\sum_{n=1}^{\infty }\left(
-1\right)
^{n}\frac{\left( 3it\right) ^{n}}{2^n\Gamma \left( n+1\right) }%
\cos x \ .
\end{equation}
The solution (121) can be written in terms of Mittag-Leffler
function for $\alpha=1$ as
\begin{equation}
\psi _{g>0}\left( x,t\right) =\cos x+\sum_{n=1}^{\infty
}\frac{\left( -\left( t/\tau \right)\right) ^{n}}{\Gamma \left( n+1
\right) }\cos x=E_{1 }\left( -\left( t/\tau \right)\right)\cos x
\end{equation}
where we set  $3i/2=1/\tau$. Eq.\,(122) can be given in exponential
form as
\begin{equation}
\psi _{g>0}\left( x,t\right) =E_{1 }\left( -\left( t/\tau
\right)\right)\cos x=e^{-t/\tau}\cos x \ .
\end{equation}
Our analytical result in Eq.\,(123) is the same with analytical
solution of GP equation for optical lattice  potential $V\left(
x\right) =\sin ^{2}x$ and attractive interactions $g>0$. The
numerical demonstrations of the Eq.\,(123) for $t=0$ and the
probability density of $\psi _{g>0}\left( x,t\right)$ are given is
given in Figs.\,(7a) and (7b), respectively. As it can be seen from
Fig.\,(7a) that the solution $\psi _{g>0}\left( x,t\right)$ for
$t=0$ has a Gaussian form, whereas probability density of $\psi
_{g>0}\left( x,t\right)$ has a double well shape for optical lattice
potential $V\left( x\right) =\sin ^{2}x$ and attractive interactions
$g=1>0$. As a result we note that these analytical and numerical
results confirm that obtained solution using HPM is the same with
analytical solution of GP equation of integer order for $V\left(
x\right)=\sin^{2}x$ and $g=1>0$.
\begin{figure*}[ht] \label{Figure 7}
\begin{center}
\includegraphics[width=5.5 in]{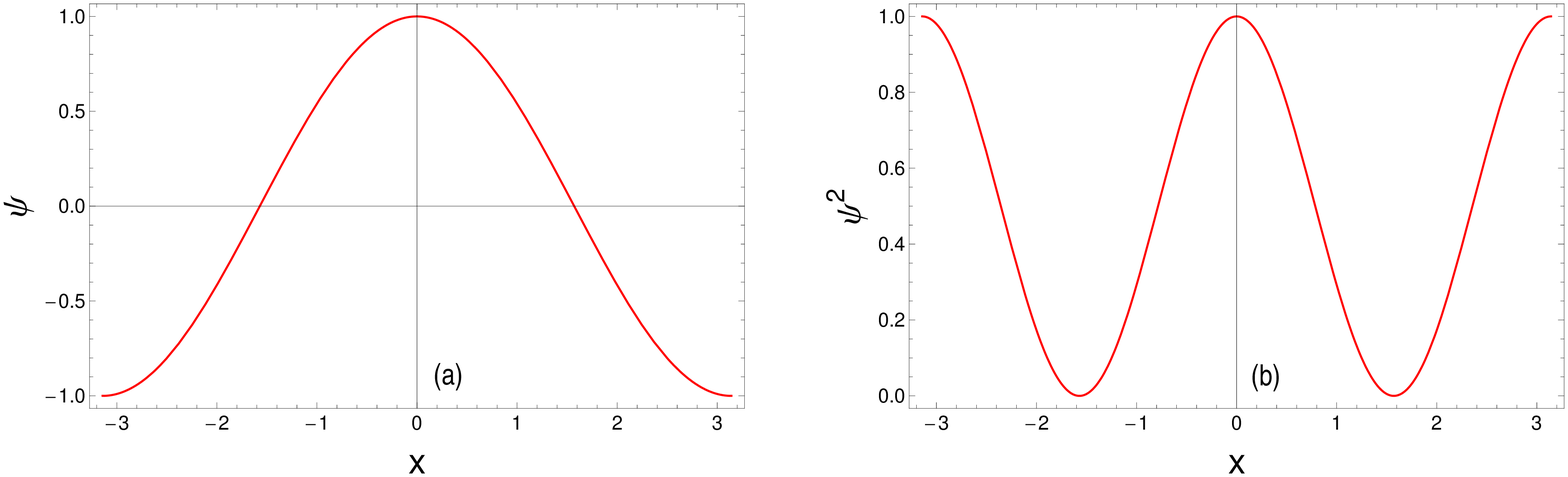}
\caption{$V\left( x\right) =\sin ^{2}x$ and repulsive interactions
$g=1>0$, a) Numerical representation of the
$\psi_{g>0}\left(x,0\right)$, b) Numerical representation of the
probability density of $\psi_{g>0}\left(x,t\right)$.}
\end{center}
\end{figure*}

\paragraph{3.Case-II} Now we consider time-fractional GP equation for
$V\left( x\right) =\sin ^{2}x$ and repulsive interaction $g=1>0$.
The iterative equation (66) of the time-fractional GP equation for
this potential is given by
\begin{equation}
\phi _{j}^{\alpha }=\frac{i}{\hbar \Gamma \left( \alpha \right) }%
\int_{0}^{t}\left( t-\tau \right) ^{\alpha -1}\left( \frac{\hbar
}{2m}\nabla ^{2}\phi _{j-1}^{\alpha }-\sin ^{2}x\phi _{j-1}^{\alpha
}-g\sum_{i=0}^{j-1}\sum_{k=0}^{j-i-1}\left\vert \phi _{i}^{\alpha
}\right\vert \left\vert \phi _{k}^{\alpha }\right\vert \phi
_{j-k-1}^{\alpha }\right) d\tau
\end{equation}
For $g=1>0$, the HPM solutions of $\phi _{j}^{\alpha }$ are given by
\begin{equation}
\phi _{0}^{\alpha }=\cos x
\end{equation}
\begin{equation}
\phi _{1}^{\alpha }=-\frac{3it^{\alpha }}{2\Gamma \left( \alpha +1\right) }%
\cos x
\end{equation}
\begin{equation}
\phi _{2}^{\alpha }=-\frac{9t^{2\alpha }}{4\Gamma \left( 2\alpha +1\right) }%
\cos x
\end{equation}
\begin{equation}
\phi _{3}^{\alpha }=\frac{27it^{3\alpha }}{8\Gamma \left( 3\alpha +1\right) }%
\cos x
\end{equation}
\begin{equation}
\begin{array}{c}
. \\
. \\
.%
\end{array}
\end{equation}
\begin{equation}
\phi _{n}^{\alpha }=\left( -1\right) ^{n}\frac{\left( 3it^{\alpha
}\right) ^{n}}{2^n\Gamma \left( n\alpha +1\right) }\cos x
\end{equation}
where we set $\hbar=m=1$ for simplicity. These terms can be put in a
seises to obtain $\psi ^{\alpha }\left( x,t\right)$
\begin{equation}
\psi ^{\alpha }\left( x,t\right) =\phi _{0}^{\alpha }\left( x\right)
+p\phi _{1}^{\alpha }\left( x\right) +p^{2}\phi _{2}^{\alpha }\left(
x\right) +...=\sum_{j=0}^{\infty }p^{j}\phi _{j}^{\alpha }\left(
x\right)
\end{equation}
We obtain final results for $g=1>0$
\begin{equation}
\psi _{g>0}^{\alpha }\left( x,t\right) =\cos x+\sum_{n=1}^{\infty
}\left( -1\right)
^{n}\frac{\left( 3it^{\alpha }\right) ^{n}}{2^n\Gamma \left( n\alpha +1\right) }%
\cos x \ .
\end{equation}
This solution can be written in terms of Mittag-Leffler function as
\begin{equation}
\psi _{g>0}^{\alpha }\left( x,t\right) =\cos x+\sum_{n=1}^{\infty
}\frac{\left( -\left( t/\tau \right) ^{\alpha }\right) ^{n}}{\Gamma
\left( n\alpha+1 \right) }\cos x=E_{\alpha }\left( -\left( t/\tau
\right) ^{\alpha }\right)\cos x
\end{equation}
where we set  $3i/2=1/\tau^\alpha$. According to Eqs.\,(31) and (32)
the asymptotic behavior of the Mittag-Leffler function, the
fractional solution (133) is given by stretched exponential form
\begin{equation}
\psi _{g>0}^{\alpha }\left( x,t\right) =E_{\alpha }\left( -\left(
t/\tau \right) ^{\alpha }\right)\cos x  \sim  \exp \left(
-\frac{\left( t/\tau \right) ^{\alpha }}{\Gamma \left( \alpha+1
\right) }\right)\cos x
\end{equation}
or for long-time regime is given by inverse power-law as
\begin{equation}
\psi _{g>0}^{\alpha }\left( x,t\right) = E_{\alpha }\left( -\left(
t/\tau \right) ^{\alpha }\right)\cos x \sim  \left( \frac{t}{\tau
}\right) ^{\alpha }\frac{1}{\Gamma \left( 1-\alpha \right) } \cos x
\ .
\end{equation}
Eqs.\,(134) and (135) are solutions of the time-fractional GP
equation for $V\left(x\right)=\sin^{2}x$ and $g=1>0$. It can be
clearly seen from Eqs.\,(134) and (135) that the time-dependent
solution of time-fractional GP equation (16) is different from
standard GP equation (1). Indeed, solutions (134) and (135) of
time-fractional GP equation indicate that for $\alpha<1$ the ground
state dynamics of the Bose-Einstein condensation evolute with time
in complex space as obey to stretched exponential for short time
regime and power law for long-time regime in the case external
potential $V\left(x\right)=\sin^{2}x$ and interactions between
bosonic particles are repulsive. Whereas for the $\alpha=1$, the
ground state dynamics of the condensation exponentially evolutes
with time. However, the spatial part of both solution in Eq.\,(133)
and in Eq.\,(123) are equal. Therefore, it is excepted that the
time-fractional dynamics of condensation also produces double well
shape behavior similar to Eq.\,(135) as it can be seen in
Fig.\,(7b). On the other hand, here we must remark that the
fractional parameter $\alpha$ is a measure of fractality in between
particles in condensation progress. Hence it determine time
evolution of the ground state dynamics of condensation.
\begin{figure*}[ht] \label{Figure 8}
\begin{center}
\subfigure[\hspace{-1.1cm}]{\label{--}
\includegraphics[width=2.9 in]{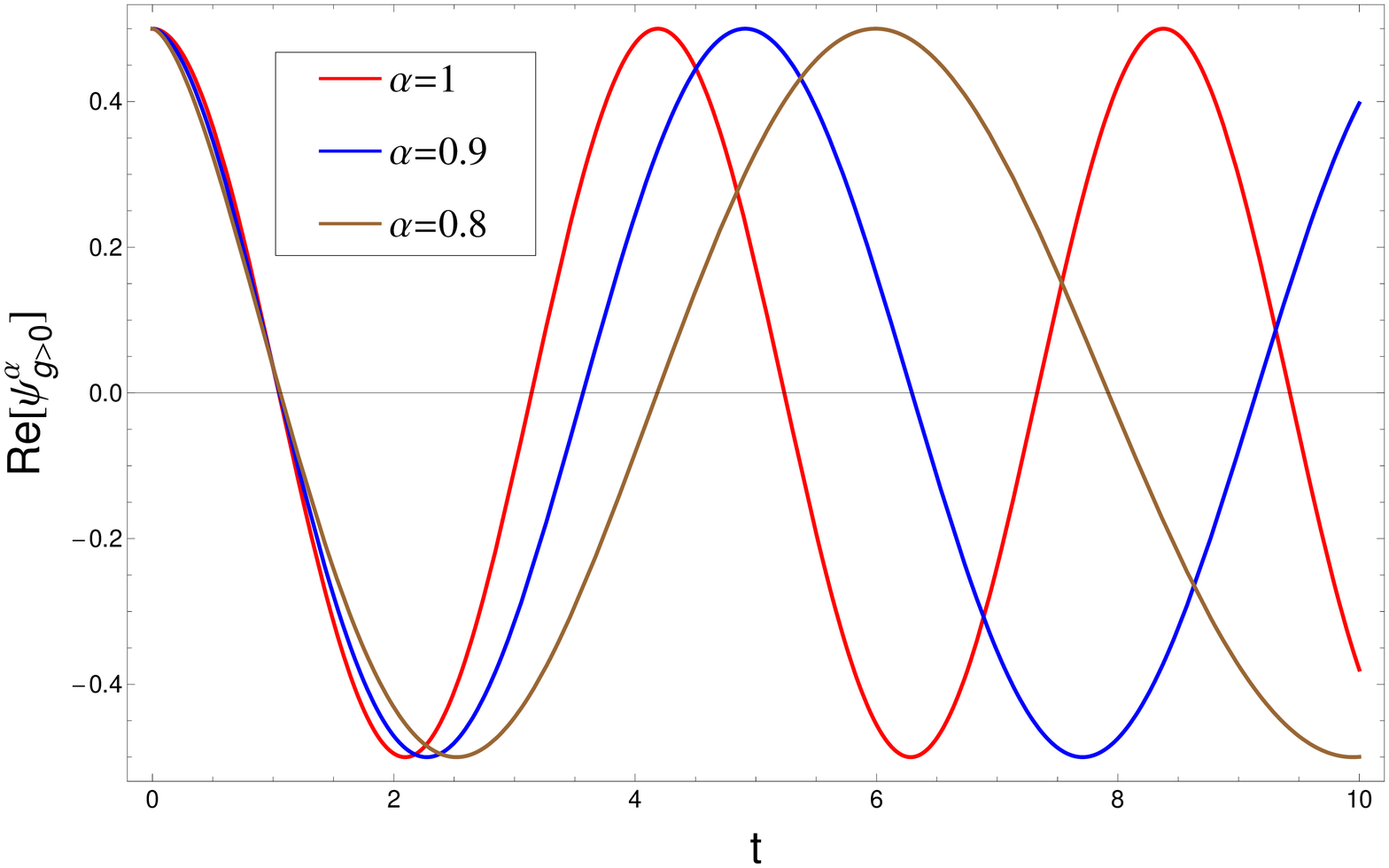}}
\hspace{0.01cm} \subfigure[\hspace{-1.1cm}]{\label{--}
\includegraphics[width=2.9 in]{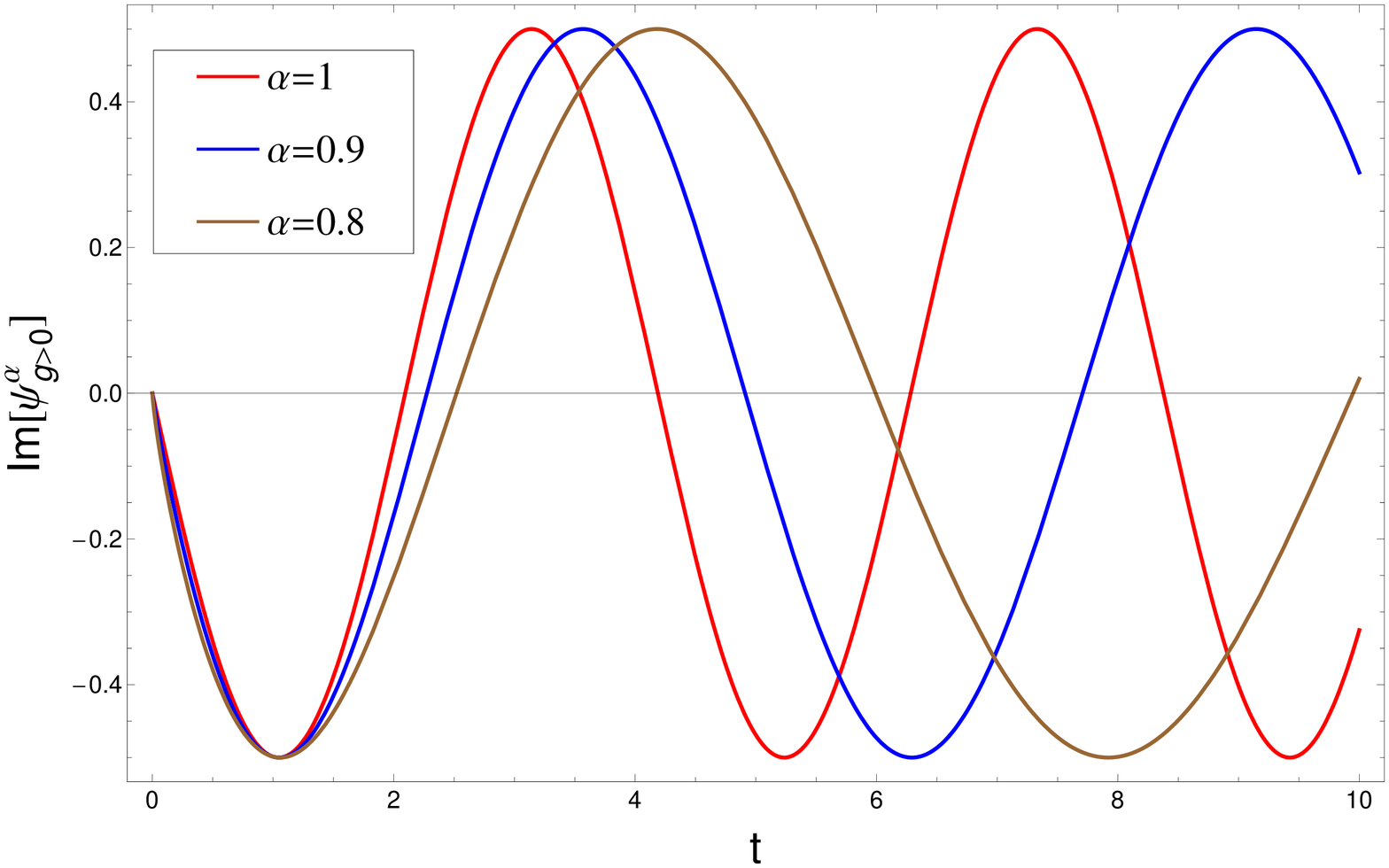}}
\hspace{0.01cm} \caption{In the case $V\left(x\right)=\sin^{2}x$ and
$g>0$ a) Real part of the $\psi^{\alpha}_{g>0}\left( x,t\right)$ for
$\alpha=1, 0.9, 0.8$ values, b) Imaginer part of the
$\psi^{\alpha}_{g>0}\left( x,t\right)$ for $\alpha=1, 0.9, 0.8$
values.}
\end{center}
\end{figure*}
In order to demonstrate effect of the fractional parameter on the
ground state dynamics or another say to investigate ground state
dynamics of condensation with fractality we plot the real and
imaginer part of the $\psi^{\alpha}_{g>0}\left( x,t\right)$ for
several $\alpha$ values.
\begin{figure*}\label{Figure 9}
\centering{
\includegraphics[width=3.5in]{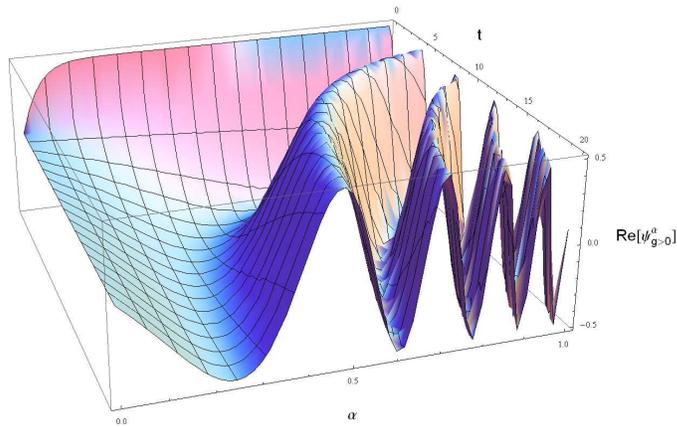}
\caption{Time and fractional parameter $\alpha$  dependence of the
real part of $\psi^{\alpha}_{g>0}\left(x,t\right)$ for
$V\left(x\right)=\sin^{2}x$ and $g>0$.}}
\end{figure*}
As it can be seen from Figure 8 that the time evolution of the real
and imaginer part of the wave function $\psi^{\alpha}_{g>0}\left(
x,t\right)$ clearly depend on fractional parameter $\alpha$. For
small values of the time, wave function solution depend on time
nearly coincident for different $\alpha$ values, however fractional
parameter $\alpha$ values substantially affect the solution of
$\psi^{\alpha}_{g>0}\left( x,t\right)$ when time is increased.
Indeed for $\alpha<1$ values this effect can be clearly seen in
Figure 8 (a) and (b) at large time values. Furthermore, to see
effect of the fractional parameter on the ground state dynamics of
condensation of bosonic particles with fractality, we give the
surface plot in Figure 9. This figure clearly shows that the real
part of the wave function $\psi^{\alpha}_{g>0}\left(x,t\right)$
evolutes to make oscillation with time for arbitrary $\alpha$
values, however, for in the limit $\alpha \rightarrow 0$, at the
same time, or in the limit $t \rightarrow 0$ the wave function
$\psi^{\alpha}_{g>0}\left( x,t\right)$ goes to stationary state
which means that the wave function in these limit does not change.
Hence in these limit values, stationary solution of the
time-fractional GP equation is independent of time.


\paragraph{\textbf{Example 4}}

Finally, we will obtain solutions of the GP equation of integer and
fractional order for optical lattice potential $V\left( x\right)
=-\sin ^{2}x$ and attractive interaction i.e. $g=-1<0$. Hence we
will consider GPE of integer order in \emph{4.Case-I} and GPE of
fractional order in \emph{4.Case-II}. We note that we have found the
analytical solution of GP equation as $\psi _{g=-1}\left( x,t\right)
=e^{it/2}\cos x$ in the case $g=-1<0$ for optical lattice potential
$V\left( x\right) =-\sin ^{2}x$ which satisfy Eq.\,(1).

\paragraph{4.Case-I} For optical lattice potential $V\left( x\right) =-\sin
^{2}x$ and attractive interaction $g=-1<0$, the iterative relation
(53) is written as
\begin{equation}
\phi _{j}=\frac{i}{\hbar }%
\int_{0}^{t}\left( \frac{\hbar }{2m}\nabla ^{2}\phi _{j-1}+\sin
^{2}x\phi _{j-1}-g\sum_{i=0}^{j-1}\sum_{k=0}^{j-i-1}\left\vert \phi
_{i}\right\vert \left\vert \phi _{k}\right\vert \phi _{j-k-1}\right)
d\tau
\end{equation}
For this potential, the HPM solutions of $\phi _{j}$ are given by
\begin{equation}
\phi _{0}=\cos x
\end{equation}
\begin{equation}
\phi _{1}=-\frac{it}{2\Gamma \left(1\right) }%
\cos x
\end{equation}
\begin{equation}
\phi _{2}=-\frac{t^2}{4\Gamma \left( 3\right) }%
\cos x
\end{equation}
\begin{equation}
\phi _{3}=-\frac{it^{3}}{8\Gamma \left( 4\right) }%
\cos x
\end{equation}
\begin{equation}
\begin{array}{c}
. \\
. \\
.%
\end{array}
\end{equation}
\begin{equation}
\phi _{n}=\left( -1\right) ^{n}\frac{\left( it\right)
^{n}}{2^n\Gamma \left( n+1\right) }\cos x
\end{equation}
where we set $\hbar=m=1$ for simplicity. If these terms are put in
the series
\begin{equation}
\psi \left( x,t\right) =\phi _{0}\left( x\right) +p\phi _{1}\left(
x\right) +p^{2}\phi _{2}\left( x\right) +...=\sum_{j=0}^{\infty
}p^{j}\phi _{j}\left( x\right)
\end{equation}
we obtain final results as
\begin{equation}
\psi _{g<0}\left( x,t\right) =\cos x+\sum_{n=1}^{\infty }\left(
-1\right)
^{n}\frac{\left( it\right) ^{n}}{2^n\Gamma \left( n+1\right) }%
\cos x \ .
\end{equation}
The solution (144) can be written in terms of Mittag-Leffler
function for $\alpha=1$ as
\begin{equation}
\psi _{g<0}\left( x,t\right) =\cos x+\sum_{n=1}^{\infty
}\frac{\left( -\left( t/\tau \right)\right) ^{n}}{\Gamma \left(
n\alpha+1 \right) }\cos x=E_{1 }\left( -\left( t/\tau
\right)\right)\cos x
\end{equation}
where we set  $i/2=1/\tau$. Eq.\,(145) can be given in exponential
form as
\begin{equation}
\psi _{g<0}\left( x,t\right) =E_{1 }\left( -\left( t/\tau
\right)\right)\cos x=e^{-t/\tau}\cos x \ .
\end{equation}
\begin{figure*}[ht] \label{Figure 10}
\begin{center}
\includegraphics[width=5.5 in]{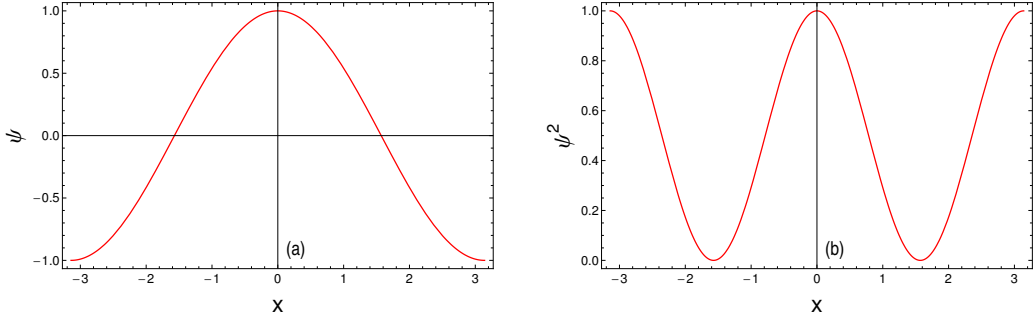}
\caption{$V\left( x\right) =-\sin ^{2}x$ and attractive interactions
$g=-1<0$, a) Numerical representation of the
$\psi_{g<0}\left(x,0\right)$, b) Numerical representation of the
probability density of $\psi_{g<0}\left(x,t\right)$.}
\end{center}
\end{figure*}
Our HPM result in Eq.\,(146) is the same with analytical solution of
GP equation for optical lattice  potential $V\left( x\right) =-\sin
^{2}x$ and attractive interactions $g<0$. At $t=0$, the numerical
demonstrations of the $\psi _{g<0}\left( x,0\right)$ and the
probability density of $\psi _{g<0}\left( x,0\right)$ are given is
given in Figs.\,(10a) and (10b), respectively. As it can be seen
from Fig.\,(7a) that the solution $\psi _{g<0}\left( x,0\right)$ for
$t=0$ has a Gaussian form, whereas probability density of $\psi
_{g<0}\left( x,t\right)$ has a double well shape for optical lattice
potential $V\left( x\right) =-\sin ^{2}x$ and attractive
interactions $g=-1<0$. As a result we note that these analytical and
numerical results confirm that obtained solution using HPM is the
same with analytical solution of GP equation of integer order for
$V\left( x\right)=-\sin^{2}x$ and $g=-1<0$.

\paragraph{4.Case-II} In last case, we will consider time-fractional GP equation
for optical lattica potential $V\left( x\right) =-\sin ^{2}x$ and
attractive interaction $g=-1<0$. The iterative equation (66) of the
time-fractional GP equation for this potential is given by
\begin{equation}
\phi _{j}^{\alpha }=\frac{i}{\hbar \Gamma \left( \alpha \right) }%
\int_{0}^{t}\left( t-\tau \right) ^{\alpha -1}\left( \frac{\hbar
}{2m}\nabla ^{2}\phi _{j-1}^{\alpha }+\sin ^{2}x\phi _{j-1}^{\alpha
}-g\sum_{i=0}^{j-1}\sum_{k=0}^{j-i-1}\left\vert \phi _{i}^{\alpha
}\right\vert \left\vert \phi _{k}^{\alpha }\right\vert \phi
_{j-k-1}^{\alpha }\right) d\tau
\end{equation}
For this potential, the HPM solutions of $\phi _{j}$ are given by
\begin{equation}
\phi _{0}^\alpha=\cos x
\end{equation}
\begin{equation}
\phi _{1}^\alpha=-\frac{it^\alpha}{2\Gamma \left(\alpha+1\right) }%
\cos x
\end{equation}
\begin{equation}
\phi _{2}^\alpha=-\frac{t^{2\alpha}}{4\Gamma \left( 2\alpha+1\right) }%
\cos x
\end{equation}
\begin{equation}
\phi _{3}^\alpha=-\frac{it^{3\alpha}}{8\Gamma \left( 3\alpha+1 \right) }%
\cos x
\end{equation}
\begin{equation}
\begin{array}{c}
. \\
. \\
.%
\end{array}
\end{equation}
\begin{equation}
\phi _{n}^\alpha=\left( -1\right) ^{n}\frac{\left( it^\alpha\right)
^{n}}{2^n\Gamma \left(n\alpha+1\right) }\cos x
\end{equation}
where we set $\hbar=m=1$ for simplicity. If these terms are put in
the series
\begin{equation}
\psi^\alpha _{g<0}\left( x,t\right) =\phi _{0}^\alpha\left( x\right)
+p\phi _{1}^\alpha\left( x\right) +p^{2}\phi _{2}^\alpha\left(
x\right) +...=\sum_{j=0}^{\infty }p^{j}\phi _{j}^\alpha\left(
x\right)
\end{equation}
we obtain final results as
\begin{equation}
\psi _{g<0}^\alpha\left( x,t\right) =\cos x+\sum_{n=1}^{\infty
}\left( -1\right)
^{n}\frac{\left( it^\alpha\right) ^{n}}{2^n\Gamma \left( n\alpha+1\right) }%
\cos x \ .
\end{equation}
This solution can be written in terms of Mittag-Leffler function as
\begin{equation}
\psi _{g<0}^\alpha\left( x,t\right) =\cos x+\sum_{n=1}^{\infty
}\frac{\left( -\left( t/\tau \right)^\alpha\right) ^{n}}{\Gamma
\left( n \alpha+1 \right) }\cos x=E_{\alpha }\left( -\left( t/\tau
\right)^\alpha\right)\cos x
\end{equation}
where we set  $i/2=1/\tau^\alpha$. According to Eqs.\,(31) and (32)
the asymptotic behavior of the Mittag-Leffler function, the
fractional solution (156) is given by stretched exponential form
\begin{equation}
\psi _{g<0}^\alpha\left( x,t\right) =E_{\alpha }\left( -\left(
t/\tau \right) ^{\alpha }\right)\cos x  \sim  \exp \left(
-\frac{\left( t/\tau \right) ^{\alpha }}{\Gamma \left( \alpha+1
\right) }\right)\cos x
\end{equation}
or for long-time regime is given by inverse power-law as
\begin{equation}
\psi _{g<0}^\alpha\left( x,t\right) = E_{\alpha }\left( -\left(
t/\tau \right) ^{\alpha }\right)\cos x \sim  \left( \frac{t}{\tau
}\right) ^{\alpha }\frac{1}{\Gamma \left( 1-\alpha \right) } \cos x
\ .
\end{equation}
Eqs.\,(157) and (158) are solutions of the time-fractional GP
equation for $V\left(x\right)=-\sin^{2}x$ and $g=-1<0$. It can be
clearly seen from Eqs.\,(157) and (158) that the time-dependent
solution of time-fractional GP equation (16) is different from
solution of the standard GP equation (1). Indeed, solutions (157)
and (158) of time-fractional GP equation indicate that the ground
state dynamics of the Bose-Einstein condensation evolute with time
in complex space as obey to stretched exponential for short time
regime and power law for long-time regime in the case external
potential $V\left(x\right)=-\sin^{2}x$ and interactions between
bosonic particles are attractive. Whereas for the $\alpha=1$, the
ground state dynamics of the condensation exponentially evolutes
with time. However, the spatial part of both solution in Eq.\,(156)
and in Eq.\,(146) are equal. Therefore, it is excepted that the
time-fractional dynamics of condensation also produces double well
shape behavior similar to Eq.\,(146) as it can be seen in
Fig.\,(10b). On the other hand, here we must remark that the
fractional parameter $\alpha$ is a measure of fractality in between
particles in condensation progress. Hence it determine time
evolution of the ground state dynamics of condensation.
\begin{figure*}[ht] \label{Figure 11}
\begin{center}
\subfigure[\hspace{-0.9cm}]{\label{--}
\includegraphics[width=2.9 in]{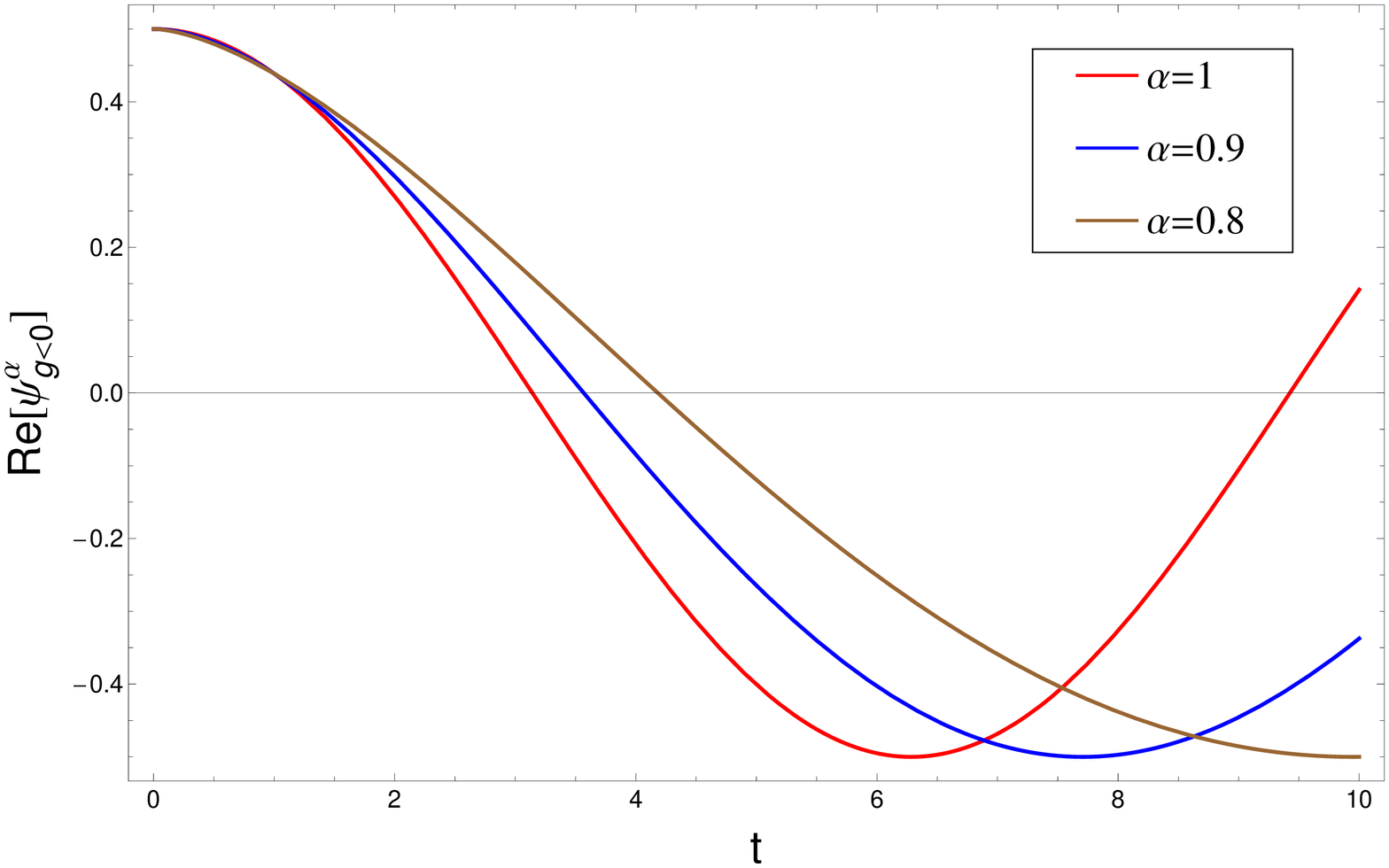}}
\hspace{0.01cm} \subfigure[\hspace{-1.1cm}]{\label{--}
\includegraphics[width=2.9 in]{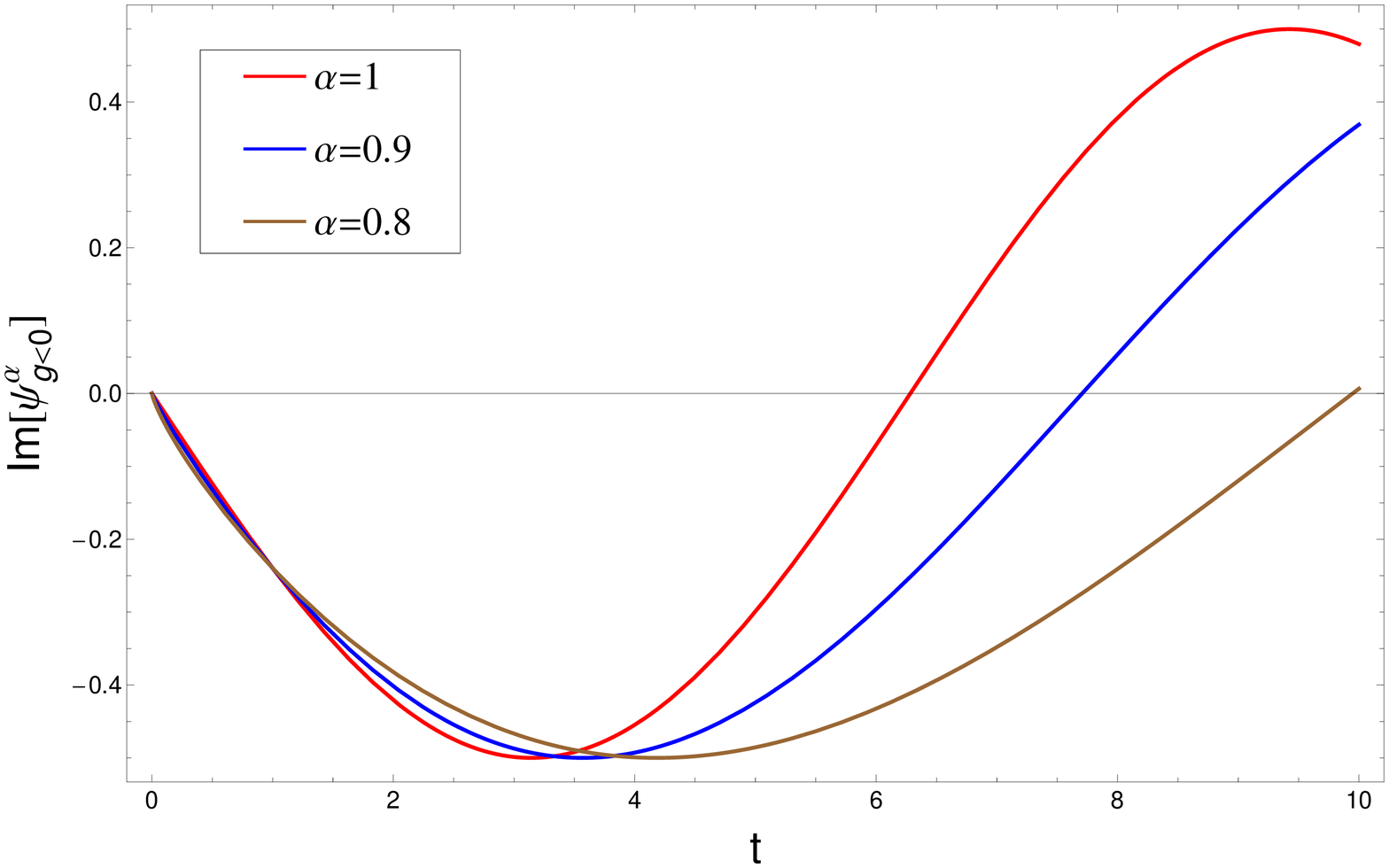}}
\hspace{0.01cm} \caption{In the case $V\left(x\right)=-\sin^{2}x$
and $g>0$ a) Real part of the $\psi^{\alpha}_{g<0}\left( x,t\right)$
for $\alpha=1, 0.9, 0.8$ values, b) Imaginer part of the
$\psi^{\alpha}_{g<0}\left( x,t\right)$ for $\alpha=1, 0.9, 0.8$
values.}
\end{center}
\end{figure*}
In order to demonstrate effect of the fractional parameter on the
ground state dynamics or another say to investigate ground state
dynamics of condensation with fractality we plot the real and
imaginer part of the $\psi^{\alpha}_{g<0}\left( x,t\right)$ for
several $\alpha$ values.
\begin{figure*}\label{Figure 12}
\centering{
\includegraphics[width=3.5in]{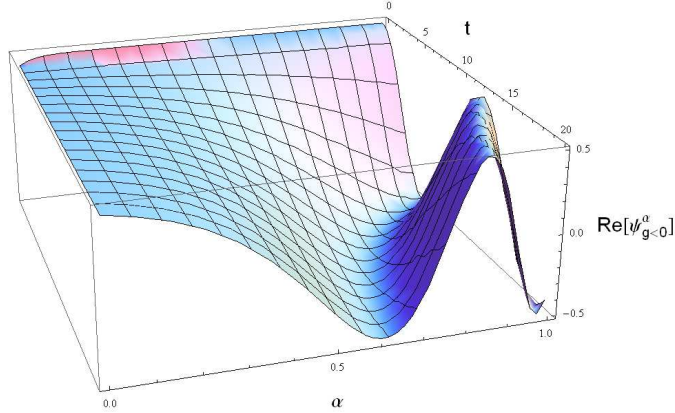}
\caption{Time and fractional parameter $\alpha$  dependence of the
real part of $\psi^{\alpha}_{g<0}\left(x,t\right)$ for
$V\left(x\right)=-\sin^{2}x$ and $g<0$.}}
\end{figure*}
As it can be seen from Figure 11 that the time evolution of the real
and imaginer part of the wave function $\psi^{\alpha}_{g<0}\left(
x,t\right)$ clearly depend on fractional parameter $\alpha$. For
small values of the time, wave function solution depend on time
nearly coincident for different $\alpha$ values, however fractional
parameter $\alpha$ values substantially affect the solution of
$\psi^{\alpha}_{g<0}\left( x,t\right)$ when time is increased.
Indeed for $\alpha<1$ values this effect can be clearly seen in
Figure 11 (a) and (b) at large time values. Furthermore, to see
effect of the fractional parameter $\alpha$ on the ground state
dynamics of condensation of bosonic particles, we give the surface
plot in Figure 12. This figure clearly shows that the real part of
the wave function $\psi^{\alpha}_{g<0}\left(x,t\right)$ evolutes to
make oscillation with time for arbitrary $\alpha$ values, however,
for in the limit $\alpha \rightarrow 0$, at the same time, or in the
limit $t \rightarrow 0$ the wave function $\psi^{\alpha}_{g<0}\left(
x,t\right)$ goes to stationary state which means that the wave
function in these limit does not change. Hence in these limit
values, stationary solution of the time-fractional GP equation is
independent of time.

\section{Conclusion}

In this study, we have suggested fractional Gross-Pitaevskii
equations to investigate the time-dependent ground state dynamics of
the Bose-Einstein condensation of weakly interacting bosonic
particle system which can includes non-Markovian processes or
non-Gaussian distributions and long-range interactions. However we
focused the time-fractional Gross-Pitaevskii equation and obtained
solutions of the the Gross-Pitaevskii equation of the integer and
fractional order for repulsive and attractive interactions in the
case external trap potentials $V(x)=0$ and optical lattice potential
$V\left( x\right) =\pm\sin^{2}x$ using Homotopy Perturbation Method.
Summarizing, in the \emph{Example 1 and 2}, we have considered GP
equation of integer and fractional order for the repulsive and
attractive interactions in the case external trap potential
$V(x)=0$. On the other hand, in the \emph{Example 3 and 4}, we have
discussed GP equation of integer and fractional order for the
repulsive and attractive interactions in the case the optical
lattice potentials $V\left( x\right) =\pm \sin^{2}x$. We found
several results: 1) we have shown that our HPM solutions of GP
equation of integer order are the same the analytical solutions of
its for $V\left(x\right)=0$ and optical lattice potentials
$V\left(x\right)=\pm \sin^{2}x$. These confirms that HPM is reliable
method to solve time-dependent non-linear partial differential
equations as well GP equation, 2) We show that the time dependent
part of the solutions of the time-fractional GP equation have
stretched exponential form for repulsive and attractive interactions
in the case $V\left(x\right)=0$ and optical lattice potentials
$V\left(x\right)=\pm \sin^{2}x$. However, for the same potentials
and interactions, time dependent part of the solution of GP equation
of integer order have exponential form. These means that time
evolution of the ground state dynamics of condensation of bosonic
particles including non-Markovian processes deviates exponential
form, and evolutes with time as stretched exponentially.

We conclude that the condensation of weakly interacting bosonic
particles have non-Markovian processes which can be lead fractional
dynamics in condensate phase. This dynamics can be represented by
time-fractional Gross-Pitaevskii equation. Although time-fractional
GP equation does not predict critical temperature $T_{c}$ of the
bosonic system, it can help to understand ground state dynamic of
these system. To understand the contradictions of the experimental
and theoretical predictions of critical temperature about
condensation in real bosonic gas, the memory effects of the
non-Markovian processes, non-Gaussian distribution of particles, the
long-range inter-particle interaction effects or the fractal
structure of the interacting in a bosonic system below the $T_{c}$
can be taken into account. In the next study, to investigate of the
condensation dynamics of weakly interacting bosonic systems we will
consider the time and space-fractional GP equation for many
different potentials, which have no analytical solutions. And also
we will try to determine relation between condensation temperature
of real bosonic system and $\alpha$ used in the fractional
Gross-Pitaevskii equation.

\section*{Acknowledgment}
This work is partially supported by Istanbul University (Project
Numbers: 12941 and 6942).

\end{document}